\documentclass[12pt,preprint]{aastex}
\usepackage{natbib,lscape}
\begin{document}

\title{Spatial Structure and Collisionless Electron Heating in Balmer-dominated Shocks}

\author{Matthew van Adelsberg\altaffilmark{1},
Kevin Heng\altaffilmark{2},
Richard McCray\altaffilmark{1},
John C. Raymond\altaffilmark{3}}

\altaffiltext{1}{JILA, University of Colorado, 440 UCB, Boulder, CO 80309; mvanadel@jilau1.colorado.edu}

\altaffiltext{2}{Institute for Advanced Study, Einstein Drive, Princeton, NJ 08540}

\altaffiltext{3}{Harvard-Smithsonian Center for Astrophysics, 60 Garden Street, Cambridge, MA 02138}

\begin{abstract}
Balmer-dominated shocks in supernova remnants (SNRs) produce strong hydrogen lines with a two-component profile 
composed of a narrow contribution from cold upstream hydrogen atoms, and a broad contribution from 
hydrogen atoms that have undergone charge transfer reactions with hot protons.  Observations of emission 
lines from edge-wise shocks in SNRs can constrain the gas velocity and collisionless electron heating at the 
shock front.  Downstream hydrogen atoms engage in charge transfer, excitation and ionization reactions, 
defining an interaction region called the shock transition zone. 
The properties of hot hydrogen atoms produced by charge transfers (called broad neutrals) are critical for 
accurately calculating the structure and radiation from the shock transition zone.  
This paper is the third in a series describing 
the kinetic, fluid and emission properties of Balmer-dominated shocks, and is the first to properly 
treat the effect of broad neutral kinetics on shock transition zone structure.  We use our models to extract shock 
parameters from observations of Balmer-dominated SNRs.  We find that inferred shock velocities and 
electron temperatures are lower than those of previous calculations by $<10\%$ for $v_s<1500$ km s$^{-1}$, and 
by $10-30\%$ for $v_s>1500$ km s$^{-1}$.  
This effect is primarily due to the fact that excitation by proton collisions and charge transfer 
to excited levels favor the high speed part of the neutral hydrogen velocity distribution.
Our results have a strong dependence on the 
ratio of electron to proton temperatures, $\beta\equiv T_e/T_p$, which allows us to construct a relation 
$\beta(v_s)$ between the temperature ratio and shock velocity.  We compare our calculations to previous results by  
\citet[][]{Ghavamianetal07b}.
\end{abstract}

\keywords{shock waves --- supernova remnants}

\section{Introduction}
\label{sect:Introduction}

Balmer-dominated shocks in supernova remnants (SNRs) encounter upstream gas 
containing a
substantial fraction of neutral hydrogen atoms.  Emission from the shock
is characterized by 
strong Balmer and Lyman lines with two-component profiles, which consist of a
narrow contribution from direct excitation of the hydrogen atoms 
entering the shock, and a broad contribution from excitation of 
``broad neutrals,'' hydrogen atoms that have been produced by charge transfer reactions with protons in 
the downstream gas \citep[][]{ChevalierRaymond78a,Chevalieretal80a}.  
Such shocks are seen in many SNRs, including parts of the 
Cygnus Loop (\citealt[][]{Ghavamianetal01a}, hereafter G01), 
Tycho and Kepler's remnants (\citealt[][]{Kirshneretal87a}, KCW87; \citealt{Smithetal91a}, S91; 
G01; \citealt[][]{Fesenetal89a}, F89), RCW 86 (G01; \citealt{Ghavamianetal07b}, G07b), 
SN 1006 (KCW87; S91; \citealt[][]{Ghavamianetal02a}, G02), and several remnants located in the 
LMC (\citealt[][]{Tuohyetal82a}, T82; S91; \citealt[][]{Ghavamianetal03a}, G03; \citealt{Ghavamianetal07a}, G07a).

Profiles of broad emission lines from edge-wise observations of shocks in 
SNRs can be used to infer shock velocities.  These calculations can be compared to 
analyses of shock front proper motion to compute distances to 
the objects \citep[][]{Chevalieretal80a,Kirshneretal87a}.  Observations 
which resolve the spatial profile of the combined H$\alpha$ emission 
can constrain the neutral fraction and density of the upstream gas \citep[][]{Raymondetal07a}.

In addition to diagnosing parameters of SNRs, Balmer-dominated shocks 
provide an important probe into the plasma physics of collisionless, 
non-relativistic shocks.  Models of the broad component width 
and integrated broad-to-narrow intensity ratio can be used 
to derive the ratio of electron to proton temperature as a function of shock 
velocity.  Such a relation can provide insight into the physical mechanisms at work in the collisionless plasma.

This is the third in a series of papers investigating the hydrodynamics, 
kinetics, and line emission from Balmer-dominated shocks.  In  
\citet[][Paper 1]{HengMcCray07a}, we calculated velocity distribution functions 
for the broad neutrals
and computed the ratio of broad-to-narrow line emission as a function 
of shock velocity.  In \citet[][Paper 2]{Hengetal07a}  
we calculated the density structure of the shock transition zone, 
where hydrogen atoms passing through the shock front undergo 
charge transfer reactions, emit radiation, and become fully ionized.

The approximations employed in Paper 2 limit the validity of the results 
to shocks entering the upstream gas with velocities $<3000$~km~s$^{-1}$, 
due to our treatment of the broad neutrals as a single fluid with the same 
bulk velocity as the ions (the ``restricted three-component model'' of Paper 2, \S3).
In fact, the broad neutrals  
in the shock transition zone are not a fluid, but have distinct anisotropic 
distribution functions depending on the number of charge transfer reactions they 
engage in (Paper 1).  In the present paper, we treat charged species in 
the shocked gas as fluids, and describe the hydrogen atoms with appropriate 
kinetic distribution functions.  
In addition, we provide an improved calculation of the broad line velocity 
profile for $v_s\ga 2000$ km s$^{-1}$.
Using this methodology, we calculate the structure of the shock 
transition zone more accurately than in Paper 2, and characterize hydrogen line emission from shocks 
with $300 < v_s < 10,000$~km~s$^{-1}$.  

We find that we can self-consistently 
determine shock parameters for most Balmer-dominated SNRs.
Our results yield lower values of the inferred shock velocity and proton-electron 
temperature equilibration than those derived using previous models. 
We compute the dependence of the electron temperature on shock velocity and compare it to 
the results of \citet[][]{Ghavamianetal07b}.
We note that our calculations are unable to fit the observations for several SNRs;  
in these cases, our basic model must be augmented by new physics to account for the data.

In \S\ref{sect:PhysicalModel} of this paper, we 
describe our physical model.  In 
\S\ref{sect:SpatialStructure}, we display the equations 
employed to calculate the structure of the shock transition zone and describe 
our numerical solution.
In \S\ref{sect:EmissionPhysics}, we compute the spatial emission profile and hydrogen line spectra
from the shock transition zone.
In \S\ref{sect:Results}, we analyze our results and describe how observations of Balmer-dominated SNRs 
should be interpreted in light of our new calculations.  Finally, in 
\S\ref{sect:Discussion}, we discuss implications of our results for collisionless electron heating, 
explore limitations of our model, and identify areas for future research.

\section{Physical Model}
\label{sect:PhysicalModel}

We consider a shock with velocity $300 < v_s < 10,000$ km s$^{-1}$ traveling 
through the ISM, which consists of cold, partially neutral hydrogen and neutral 
helium.  The pre-shock fraction of helium, relative to hydrogen, is denoted $f_{He}$.  If the total upstream 
density of protons and hydrogen atoms is $n_0$, then the fraction of pre-shock protons is 
defined to be $f_p \equiv n_p/n_0$.

In the frame of the shock, the cold upstream neutrals and charged particles flow downstream with 
uniform velocity $v_s$.  At the shock, we assume that the protons are heated in a thermal distribution to 
$T_p \sim {3m_p v_s^2 \over 16 k_B} \sim 10^7$ K for $v_s\sim 10^8$ cm s$^{-1}$.
At the high post-shock temperatures in Balmer-dominated remnants, emission from hydrogen excitation 
is much stronger than forbidden transitions in metals seen in radiative shocks \citep[e.g.,][]{ShullMcKee79a}.  
The length scale for thermalization of protons is set by the proton cyclotron gyroradius, 
$l_{\rm gyro} \sim 10^8$ cm at $B\sim 10^{-4}$ G, which is much less than the length scale for the shock transition 
zone, $l_{\rm zone} \gtrsim 1/(n_0\sigma) = 10^{14}$~cm (set by the mean-free paths for excitation, 
charge transfer and ionization reactions, and the total upstream density).

Without any energy transfer between shocked protons and electrons, the electron temperature 
is expected to be a factor of $m_e/m_p$ less than that 
of the protons.  Since the collisional timescale for temperature equilibration between electrons and protons 
is in some cases longer than the age of the remnant \citep[e.g.,][]{Spitzer62a}, we assume 
that the electrons remain at a fixed temperature 
with respect to the protons in the downstream material (note, however, that intra-species and inter-ion
collisional timescales can be much shorter).\footnote{Some collisional heating of electrons by protons will 
occur.  We estimate that such an effect can lower the derived value of $\beta$ by $1-2\%$}
However, several authors have argued that electrostatic 
instabilities at the shock front can increase the electron temperature, with estimates ranging from 
$T_e\sim 0.1T_p$ to $T_e=T_p$
\citep[see][and the references therein]{CargillPapadopoulos88a,Ghavamianetal07b}.  
Since the physics of non-relativistic, collisionless shocks is 
still poorly understood, we parametrize the ratio of electron to proton temperatures using the definition
\begin{eqnarray}
\label{eq:betaDef}
\beta \equiv {T_e\over T_p}.
\end{eqnarray}
One of the goals of our work is to use observations of Balmer-dominated remnants to constrain the value of $\beta$ 
in the shock transition zone as a function of shock velocity.

In the uniform velocity upstream gas, we assume that negligible interactions take place between neutral 
and charged species.  
The cold neutral hydrogen atoms passing through the shock are not affected by the discontinuity.  
In contrast, the bulk 
velocities of the downstream thermal protons and electrons become approximately $v_s/4$,
at which point ionization and charge transfer reactions between cold atoms and hot protons occur 
(see \S\ref{subsect:SpatialStructureResults}).  However, if a significant amount of the energy dissipated 
in the shock is used to produce cosmic rays, the ion compression ratio increases, and their bulk 
velocity decreases.  Furthermore, the presence of a cosmic ray precursor can broaden the distribution of 
upstream neutrals.  We discuss the implications of these physical effects in \S\ref{sect:Discussion}.

Downstream charge transfer reactions between cold neutrals and hot protons produce a new population of 
hot atoms, referred to as 
broad neutrals.  Initially, such an interaction will produce a cold proton, which is re-energized through 
gyro motion around the magnetic field and intra-species Coloumb collisions with the hot proton population.  
We assume that such protons rapidly equilibrate back into the thermal pool.
However, recent calculations by \citet[][]{Raymondetal08a} indicate that interactions within the shock front 
may produce a distinct ``pickup'' ion population analogous to that in the solar wind.

H$\alpha$ emission is produced by excitation reactions and charge transfers to 
excited states 
in the shock transition zone.  The narrow and broad components of the line are emitted by the cold 
hydrogen and 
broad neutrals, respectively.
The spatial and velocity distributions of the broad neutrals are critical for calculating the correct 
structure and emission from the transition zone.  In Paper 2, we assumed 
that the broad neutrals could be characterized as a single fluid, with velocity and temperature equal to that of the 
ions.  This treatment is flawed because, unlike the ionic species, the broad neutrals have negligible interactions 
among themselves.
Therefore, each time a neutral engages in charge exchange, it becomes part of a new distribution with distinct 
velocity and temperature (see Paper 1) and is not subsumed into the original population. 
This fact has not, until now, been properly taken into account in the literature.

The detailed kinetic properties of the particle distributions and their reaction rates were calculated in Paper 1.  
We use these methods, with updated values for the atomic cross sections (see Appendix~\ref{sect:AppendixA}), 
in our fluid calculation of the shock transition zone structure.  In this picture, the 
broad neutrals form an infinite set of atomic populations with distinct reaction rates.  In practice, as noted in 
Paper 1, the velocity distribution function of atoms experiencing many charge transfers rapidly converges to that of 
the protons, usually after only three or more interactions.  Thus, only three distinct broad 
components interact with the other species.  The bulk velocities and temperatures of the broad components are
written:
\begin{eqnarray}
\label{eq:vtrel1}
v_k & = & F_{vk} v_p,\\
\label{eq:Trel1}
T_k & = & F_{Tk} T_p,
\end{eqnarray}
where $v_k$ and $T_k$ are the velocity and temperature of the broad neutral population after $k$ charge transfers.  
The coefficients $F_{vk}(v_s,\beta,f_p,f_{He})$ and $F_{Tk}(v_s,\beta,f_p,f_{He})$ are calculated using the 
methods described in Paper 1 and, in principle, are also functions of $v_p$ and $T_p$.  However, we show in 
\S\ref{subsect:SpatialStructureResults} that 
the proton velocity and temperature do not vary significantly enough in the shock transition zone to affect the values 
of these coefficients.  For $k\ge 3$, $F_{vk}$ and $F_{Tk}$ are taken to be unity.

\begin{figure}
\begin{center}
\plotone{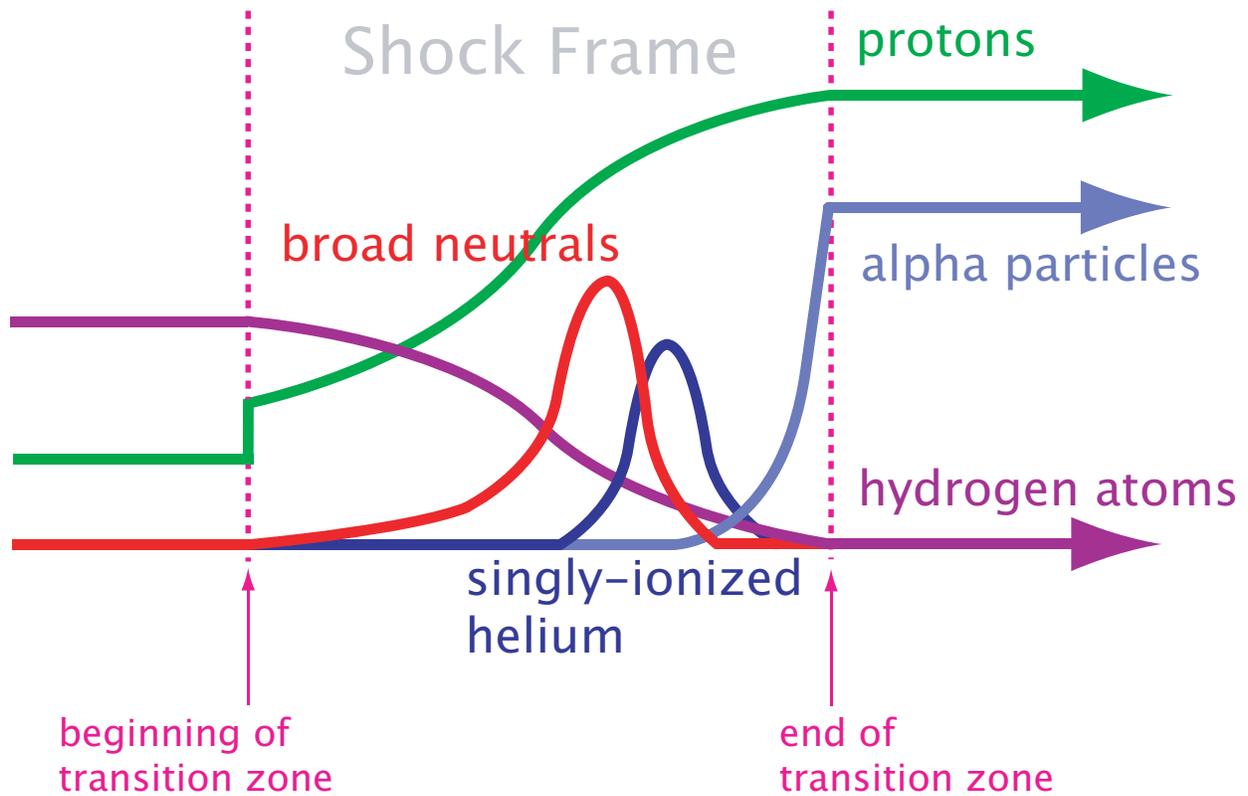}
\end{center}
\caption{Schematic depiction of density variation in the shock transition zone.  The vertical axis represents 
dimensionless 
density for the different particles species, while the horizontal axis represents distance.  The shock front is marked 
by the left dotted vertical line.  Upstream of the shock, we assume that negligible interactions take place between 
the particles (though see the discussion in \S\ref{sect:PhysicalModel}).  
After protons and electrons are isotropized and decelerated at the shock front, 
ionization and charge transfer 
reactions deplete the cold neutrals and produce hot broad atoms.  The transition 
zone terminates where all the neutral species are ionized, leaving a mix of protons, electrons and alpha particles.}
\label{fig:cartoon}
\end{figure}

Ionization of helium in the shock transition zone produces singly-ionized He$^+$ and alpha particles.  We 
assume that the helium ions and protons are coupled through rapid inter-ion Coloumb collisions.  In this case, all 
of the ions share the same bulk velocity ($v_i$) and temperature ($T_i$).
Figure~\ref{fig:cartoon} schematically displays the density variation of the different particle species in the 
shock transition zone.   

\section{Spatial Structure of the Transition Zone}
\label{sect:SpatialStructure}

\subsection{Basic Equations}
\label{subsect:BasicEquations}

We employ a plane-parallel coordinate system in the frame of the shock, in which the shock front is at $z=0$.  
The density structure of the transition zone is determined by conservation of mass flux:
\begin{eqnarray}
\label{eq:MassFlux1}
{d\over dz}\left(n_H v_s\right)  & = & -n_H \sum_{s=e}^{\alpha} n_s \tilde{R}_{iH,s} - n_H n_p \tilde{R}_{T_0,p},\\
{d\over dz}\left(n_1 v_1\right) & = & n_H n_p \tilde{R}_{T_0,p} - 
	n_1 \sum_{s=e}^{\alpha} n_s \tilde{R}_{iB,s} - n_1 n_p \tilde{R}_{T,p},\\
{d\over dz}\left(n_2 v_2\right) & = & n_1 n_p \tilde{R}_{T,p} - n_2\sum_{s=e}^{\alpha} n_s \tilde{R}_{iB,s} - 
	n_2 n_p \tilde{R}_{T,p},\\
\vdots & & \nonumber\\
\label{eq:massk}
{d\over dz}\left(n_k v_k\right) & = & n_{k-1} n_p \tilde{R}_{T,p} - n_k\sum_{s=e}^{\alpha} n_s \tilde{R}_{iB,s} - 
	n_k n_p \tilde{R}_{T,p},\\
\vdots & & \nonumber\\
{d\over dz}\left(n_{He} v_s\right) & = & -n_{He} \sum_{s=e}^{\alpha} n_s \tilde{R}_{iHe,s},\\
{d\over dz}\left(n_{He^+} v_i\right) & = & n_{He} \sum_{s=e}^{\alpha} n_s \tilde{R}_{iHe,s} - 
	n_{He^+} \sum_{s=e}^{\alpha} n_s \tilde{R}_{iHe^+,s},\\
{d\over dz}\left(n_{\alpha} v_i\right) & = & n_{He^+} \sum_{s=e}^{\alpha} n_s \tilde{R}_{iHe^+,s},\\
\label{eq:MassFlux6}
{d\over dz}\left(n_p v_i\right) & = & \sum_{s=e}^{\alpha} n_s \left(n_H \tilde{R}_{iH,s}+
	\sum_{k=1}^N n_k \tilde{R}_{iB,s}\right), 
\end{eqnarray}
where the rate coefficients $\tilde{R}$ for the atomic interactions (in units of cm$^3$ s$^{-1}$) 
are defined in Appendix~\ref{sect:AppendixA}.
Charge conservation requires the electron density to obey the relation $n_e = n_p+n_{He^+}+2n_{\alpha}$.
The subscripts 
$iH$, $iHe$, and $iHe^+$ denote ionization of hydrogen, helium, and singly-ionized helium, while the index $s$ runs 
over the charged particles participating in the reactions, 
$s=\{e,p,He^+,\alpha\}$ (electrons, protons, singly-ionized helium and alpha particles).  The subscript $T_0$ indicates 
charge transfer of cold hydrogen atoms with protons.  
As discussed in \S\ref{sect:PhysicalModel}, the subscript $k$ denotes the neutral hydrogen population which has engaged 
in $k$ charge transfers, where the index runs from $k=1$ to $k=\infty$.
While each broad population will have a separate rate 
coefficient for every atomic 
interaction, the values are not sensitive to the details of the broad distribution functions.  Therefore, 
we use one rate coefficient to describe the interactions of all the broad populations: 
the subscripts $iB$ and $T$ denote ionization and  
charge transfer, respectively. We list the references for our interaction cross sections and rate coefficients in 
Appendix~\ref{sect:AppendixA}.

The system of differential equations is completed by expressions for conservation of momentum and energy flux:
\begin{eqnarray}
\label{eq:MomFlux1}
{d\over dz}\left[m_p n_H v_s^2 +\sum_{k=1}^{\infty}\left(P_k + m_p n_k v_k^2\right) + 4 m_p n_{He} v_s^2  + 
	\sum_{s=e}^{\alpha}\left(P_s+m_s n_s v_i^2\right)\right] & = & 0,\\
\label{eq:EnergyFlux1}
{d\over dz}\left[{1\over 2} m_p n_H v_s^3 +\sum_{k=1}^{\infty}\left(P_k+U_k+ {1\over 2} m_p n_k v_k^2\right) v_k +
	\hspace{1in} \right. & & \nonumber\\ \left. 2 n_{He} v_s^3 +
	\sum_{s=e}^{\alpha}\left(P_s+U_s+{1\over 2} m_s n_s v_i^2\right) v_i\right] & = & 0,
\end{eqnarray}
where $P=n k_B T$ is the pressure and $U={3\over 2} n k_B T$ is the energy density.  The ratios $P_k/n_k$ and 
$U_k/n_k$ are set by the kinetic calculation of the broad neutral distribution functions through 
eq.~(\ref{eq:Trel1}), which assumes $T_i=3m_p v_s^2/16(1+\beta)k_B$.  
Though the ion temperature varies by as much as 10\% in 
the shock transition zone, this produces a negligible effect on the rate coefficients.
Thus, the value of $T_k$ is proportional to the local ion temperature.  Furthermore, our steady-state fluid calculation 
neglects kinetic evolution of the broad neutral distributions, which we expect to be a modest effect.

We define a natural length scale for the transition zone, according to the formula: 
\begin{eqnarray}
\label{eq:lscale}
l_{\rm z} \equiv {v_s\over n_0 \bar{R}},
\end{eqnarray}
where $\bar{R}$ is a typical value for the interaction rate coefficents, taken to be $10^{-8}$ cm$^3$ s$^{-1}$.  We 
also define a set of dimensionless variables with $\chi\equiv z/l_{\rm z}$, $\eta\equiv n/n_0$, $u\equiv v/v_s$, and 
$\epsilon\equiv k_B T/(m_p v_s^2)$.  As discussed above, broad neutrals which have engaged in three or more charge 
transfers can be described with the thermal proton distribution function.  We therefore define 
$\eta_N = \sum_{k=3}^{\infty} \eta_k$.  Summing over eqs.~(\ref{eq:massk}) with $k\ge 3$ and using these definitions, 
the formulas for the shock transition zone structure can be written:
\begin{eqnarray}
\label{eq:FinalMass1}
{d\over d\chi}\eta_H & = & -\eta_H\sum_{s=e}^{\alpha} \eta_s \mathcal{R}_{iH,s}-\eta_H \eta_p \mathcal{R}_{T_0,p},\\
{d\over d\chi}\left(\eta_1 u_1\right) & = & \eta_H \eta_p \mathcal{R}_{T_0,p} - 
	\eta_1 \sum_{s=e}^{\alpha} \eta_s \mathcal{R}_{iB,s} - \eta_1 \eta_p \mathcal{R}_{T,p},\\
{d\over d\chi}\left(\eta_2 u_2\right) & = & \eta_1 \eta_p \mathcal{R}_{T,p} - 
	\eta_2 \sum_{s=e}^{\alpha} \eta_s \mathcal{R}_{iB,s} - \eta_2 \eta_p \mathcal{R}_{T,p},\\
{d\over d\chi}\left(\eta_N u_i\right) & = & \eta_2 \eta_p \mathcal{R}_{T,p} - 
	\eta_N \sum_{s=e}^{\alpha} \eta_s \mathcal{R}_{iB,s},\\
{d\over d\chi}\eta_{He} & = & -\eta_{He} \sum_{s=e}^{\alpha} \eta_s \mathcal{R}_{iHe,s},\\
{d\over d\chi}\left(\eta_{He^+} u_i\right) & = & \eta_{He}\sum_{s=e}^{\alpha} \eta_s \mathcal{R}_{iHe,s} 
	- \eta_{He^+} \sum_{s=e}^{\alpha} \eta_s \mathcal{R}_{iHe^+,s},\\
{d\over d\chi}\left(\eta_{\alpha} u_i\right) & = & \eta_{He^+}\sum_{s=e}^{\alpha} \eta_s \mathcal{R}_{iHe^+,s},\\
\label{eq:FinalMass8}
{d\over d\chi}\left(\eta_p u_i\right) & = & \sum_{s=e}^{\alpha} \eta_s \left(\eta_H\mathcal{R}_{iH,s} + 
	\sum_{k=1}^N \eta_k \mathcal{R}_{iB,s}\right),
\end{eqnarray}
\begin{eqnarray}
\label{eq:FinalMom}
{d\over d\chi}\left[\eta_H+\sum_{k=1}^N\left(\eta_k\epsilon_k+\eta_k u_k^2\right) + 4 \eta_{He} + 
	\sum_{s=e}^{\alpha}\left(\eta_s\epsilon_s + {m_s\over m_p} \eta_s u_i^2\right)\right] & = & 0,\\
\label{eq:FinalEnergy}
{d\over d\chi}\left[\eta_H+\sum_{k=1}^N\left(5\eta_k\epsilon_k u_k + \eta_k u_k^3\right) + 4 \eta_{He} +
	\sum_{s=e}^{\alpha}\left(5\eta_s\epsilon_s u_i + {m_s\over m_p} \eta_s u_i^3\right)\right] & = & 0,
\end{eqnarray}
where $\mathcal{R}\equiv \tilde{R}/\bar{R}$.

\subsection{Solution Method}
\label{subsect:SolutionMethod}

Eqs.~(\ref{eq:FinalMass1})--(\ref{eq:FinalEnergy}) are a set of ten coupled, non-linear ordinary differential equations, 
which are solved with the constraint of charge conservation.  We compute eqs.~(\ref{eq:FinalMass1})--(\ref{eq:FinalMass8}) 
directly in terms of the number density flux, denoted $y \equiv \eta u$.
The ion velocity and temperature, $u_i$ and $\epsilon_i$, determine the broad velocity $u_k$ and temperature $\epsilon_k$ 
through the relations (\ref{eq:vtrel1})-(\ref{eq:Trel1}).  Once $y$ and $u$ 
are known, the dimensionless number density $\eta = y/u$ can be substituted into the right hand side of 
the conservation equations.

We solve eqs.~(\ref{eq:FinalMom}) and (\ref{eq:FinalEnergy}) subject to the boundary conditions $y_H(0)~=~1-f_p$, 
$y_{He}(0)~=~(1-f_p)f_{He}$, $y_p(0)~=~f_p$, and $y_k(0)~=~y_{He^+}(0)~=~y_{\alpha}(0)~=~0$.  
The upstream neutral temperature is 
taken to be zero, while the protons and electrons are in thermal distributions with temperatures $T_u\ll T_p$.  
Integrating these equations, substituting 
the initial conditions, and using the definition for $y$ yields:
\begin{eqnarray}
\label{eq:quad1}
\left(\sum_{k=1}^N F_{v,k} y_k + \sum_{s=p}^{\alpha} {m_s\over m_p} y_s\right) u_i^2 + 
	\left[y_H + 4 y_{He} - 1 - (1+\beta)f_p \epsilon_u -4(1-f_p)f_{He}\right]u_i + \nonumber\\
	\left(\sum_{k=1}^N {F_{T,k}\over F_{v,k}} y_k + \beta y_e + \sum_{s=p}^{\alpha} y_s\right)\epsilon_i = 0,\\
\label{eq:quad2}
\left(\sum_{k=1}^N F_{v,k}^2 y_k + \sum_{s=p}^{\alpha} {m_s\over m_p}y_s\right)u_i^2 + 
	\left(5 \sum_{k=1}^N F_{T,k} y_k + 5\beta y_e + 5\sum_{s=p}^{\alpha} y_s\right) \epsilon_i +\nonumber\\
 	y_H + 4 y_{He} - 1 - 5(1+\beta)f_p \epsilon_u -4(1-f_p)f_{He} = 0.
\end{eqnarray}
Note that the sums over charged species run from protons to alpha particles and do not include electrons, which appear 
separately in the equations (we neglect terms proportional to $m_e/m_p$).  The 
system (\ref{eq:quad1})--(\ref{eq:quad2}) can be solved simultaneously to yield a quadratic equation in 
$u_i$ with coefficients that are functions of $\chi$.  The solution to the quadratic has two positive roots.  
Only one of the roots is less than unity; it represents the 
physical solution for the bulk ion velocity, which must be less than the shock velocity.
We use a standard Runge-Kutta method to solve eqs.~(\ref{eq:FinalMass1})--(\ref{eq:FinalEnergy}), at 
each integration step using (\ref{eq:quad1})--(\ref{eq:quad2}) to compute the ion velocity which determines the 
particle densities.

\begin{figure}
\begin{center}
\plottwo{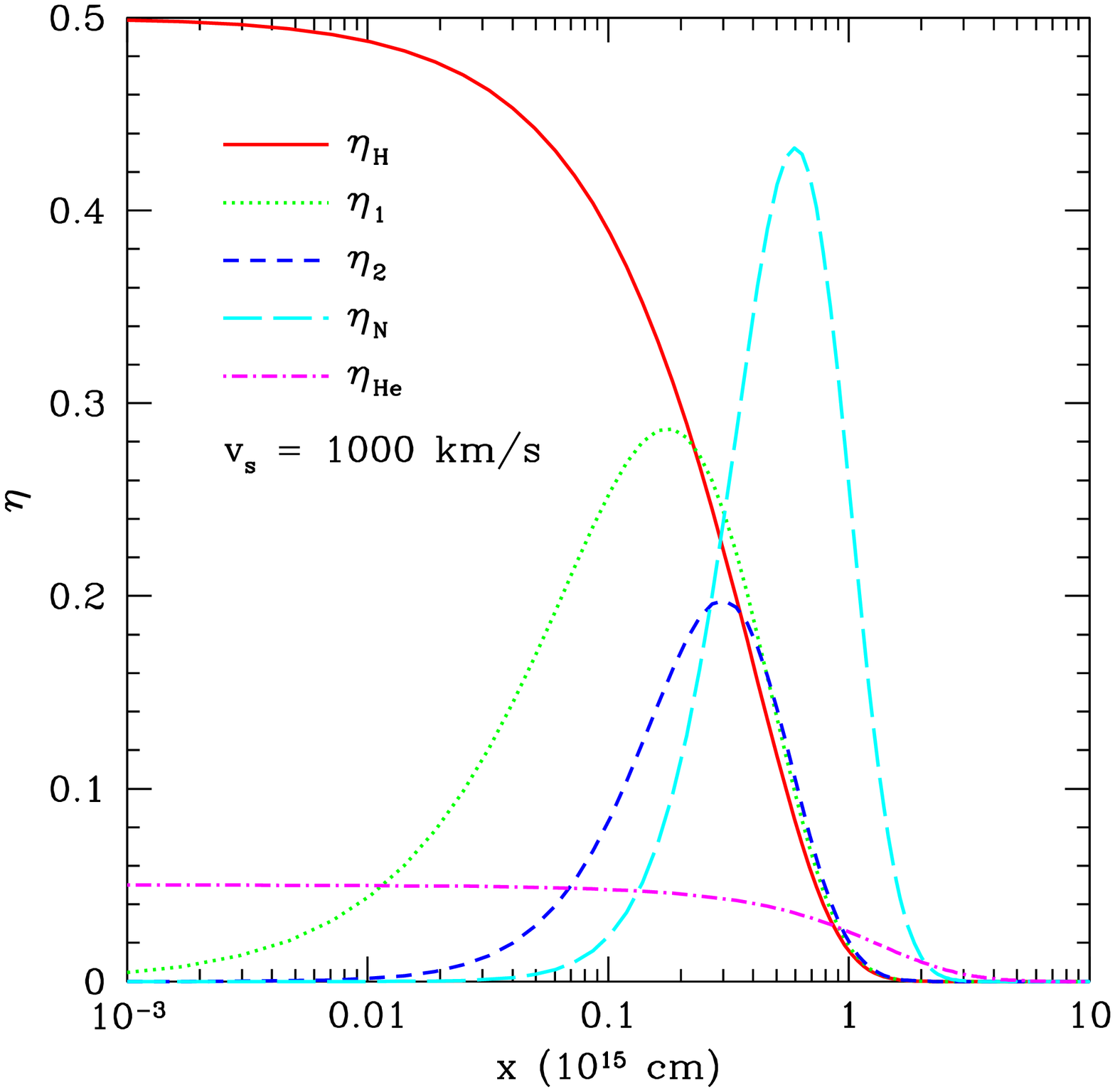}{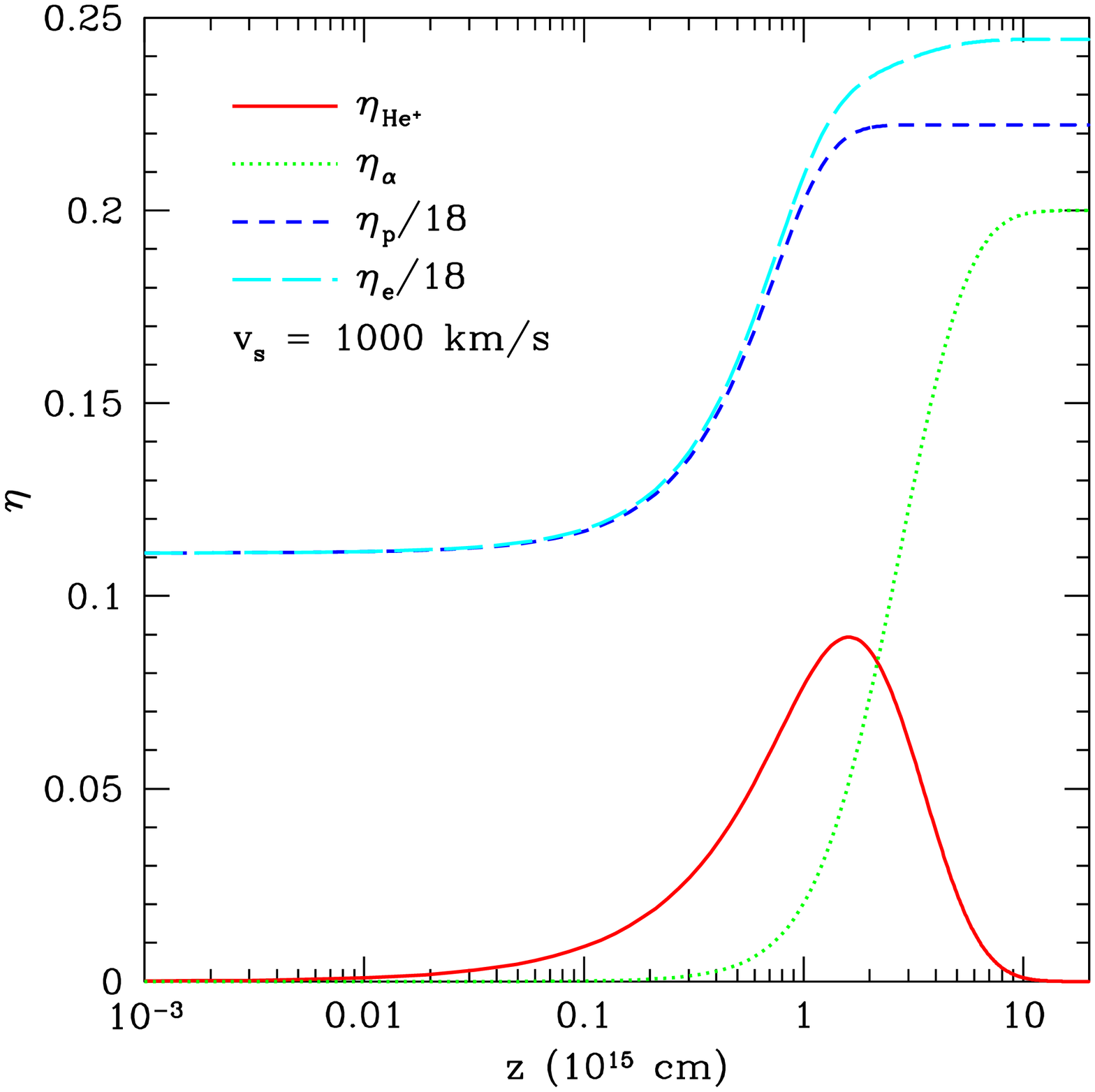}
\end{center}
\caption{Dimensionless density as a function of physical distance behind the shock front, for parameters 
$v_s=1000$ km s$^{-1}$, $f_p=0.5$, $\beta=1$, $n_0=1$ cm$^{-3}$, and $f_{He}=0.1$.  
The left panel shows the shock transition zone 
structure for the neutral species; the solid and dash-dotted curves show the density of neutral hydrogen and helium, 
respectively.  Charge transfer reactions produce broad neutrals which have undergone one (dotted curve), 
two (short-dashed) 
and three or more (long-dashed) charge transfer reactions.  
The right panel shows the transition zone structure for the charged species. 
Note that the densities of the protons and electrons have been scaled by a factor of $1/18$.  Singly-ionized helium 
(solid curve) is produced downstream from the shock and ionized to produce alpha particles (dotted curve).  The 
proton (short-dashed curve) and electron (long-dashed curve) densities saturate when the neutral species are fully 
ionized.}
\label{fig:denb1f5v1000}
\end{figure}

Figure~\ref{fig:denb1f5v1000} shows the density structure of the 
shock transition zone for $v_s=1000$ km~s$^{-1}$, $f_p=0.5$, and $\beta=1$.  
Using eq.~(\ref{eq:lscale}), the spatial coordinate 
has been converted from $\chi$ to physical distance $z$ behind the shock front. 
We use external density $n_0=1$ cm$^{-3}$ and helium fraction $f_{He}=0.1$ for all calculations 
in this paper.
The left and right panels of Figure~\ref{fig:denb1f5v1000} show the dimensionless densities for the 
neutral and charged species, 
respectively.  The electron and proton densities have been scaled by a factor of $1/18$.  
In the left panel, the solid curve shows a monotonic decrease in the density of 
the cold hydrogen atoms, which are removed by both charge transfer and ionization reactions.  Charge transfer 
produces three populations of broad neutrals.  At low velocities, charge transfer dominates over ionization, 
and many broad neutrals with $k>3$ are produced.  The dash-dotted curve shows the density of neutral helium, 
which is ionized farther downstream than hydrogen due to its smaller ionization rate coefficients.  In the right panel, 
the solid curve shows singly-ionized helium, which is produced downstream the from neutral atoms, and is then 
ionized to yield a monotonically increasing population of alpha particles, shown by the dotted curve.  
The proton and electron 
densities are depicted by the short-dashed and long-dashed curves, which saturate when all of the neutral species have 
been depleted.  The final electron density is slightly higher than that of the protons due to the presence of 
alpha particles.

\section{Line Emission}
\label{sect:EmissionPhysics}

Once the density structure of the transition zone is determined, we can compute the hydrogen line 
emission, including the spatial distribution and line profiles for the broad and narrow components.  We  
neglect collisional dexcitation and assume that every atom is excited from the ground state. 
H$\alpha$ photons are produced by transitions from atomic levels $3s$ and $3d$ to $2p$, as well 
as from $3p$ to $2s$.  
In the latter case, the atomic physics is complicated by a possible transition from $3p$ directly to 
$1s$, which results in a Lyman $\beta$ photon.  
If the medium is optically 
thin to Ly$\beta$ photons (Case A conditions), this possibility can be taken into account by proper weighting 
of the distinct angular momentum states in constructing excitation and change transfer 
cross sections for H$\alpha$ emission:
\begin{eqnarray}
\label{eq:sigHalpha}
\sigma_{H\alpha} = \sigma_{3s} + \sigma_{3d} + B_{3p,2s}\sigma_{3p}, 
\end{eqnarray}
where the factor $B_{3p,2s}\approx 0.12$ is the fraction of transitions from $3p$ to $2s$.  
If the medium is optically thick to Ly$\beta$ emission (Case B conditions), re-absorption 
by ground state hydrogen effectively traps Ly$\beta$ photons until they are re-emitted as H$\alpha$ photons.
In this case, 
all of the transitions eventually result in H$\alpha$ emission, and we set $B_{3p,2s}\approx 1$.

Case A and B conditions represent the two extremes of media that are optically thin and thick to 
Ly$\beta$ scattering.
For the stationary atoms (i.e., cold hydrogen) which produce the narrow line,  
the optical depth to scattering of Ly$\beta$ photons is, at line center, 
$\tau_{\beta}\sim n_H \sigma l_{\rm zone} \ga 1$, with $n_H\sim 0.5$ cm$^{-3}$, 
$\sigma\sim 10^{-14}$ cm$^{-2}$, and $l_{\rm zone}\sim 2\times 10^{14}$ cm 
\citep[see, e.g.,][]{RybickiLightman79a,Cox01a}.  
The column of upstream neutrals will also line scatter Ly$\beta$ emission, increasing the 
effective value of $\tau_{\beta}$.
Thus, in conditions appropriate for many Balmer-dominated SNRs, partial 
scattering of Lyman photons will occur, producing results intermediate between Cases A and B for 
the narrow line \citep[][]{Ghavamianetal01a}.  
 
The conversion efficiency of narrow component Ly$\beta$ to H$\alpha$ was first computed by 
\citet[][]{Chevalieretal80a}.  Subsequent Monte Carlo calculations were performed by \citet[][]{Lamingetal96a} and 
\citet[][]{Ghavamianetal01a}.  
Briefly, models of the neutral hydrogen density and narrow component excitation rate 
are computed as a function of distance behind the shock.
These are used to calculate the profile of excitations to the 3p 
level.  Photons are emitted at frequencies distributed according to the pre-shock temperature in random directions. 
They are followed as they are absorbed and re-emitted as Ly$\beta$ or H$\alpha$ at a different locations until they 
escape.  The conversion fraction depends on shock speed, electron-ion temperature ratio, and pre-shock ionization 
fraction, but not on total upstream density.  We assume that the variation with $f_p$ is modest, and that the temperature  
ratio decreases from $\beta\sim 1$ for $v_s\la 500$ km s$^{-1}$ to $\beta<0.1$ for $v_s>2000$ km s$^{-1}$, and use 
the calculations of \citet[][]{Lamingetal96a}, in which $60-75\%$ of 3p excitations result in Ly$\beta$ photons.  We 
fit the resulting conversion fraction as a function of shock velocity, yielding:
\begin{eqnarray}
\label{eq:fitfrac}
B_{3p,2s} = 12 v_{s,5}^{-1} + (0.63+3.6\times 10^{-5}v_{s,5}),
\end{eqnarray}
where $v_{s,5}$ is the shock velocity in units of $10^5$ cm s$^{-1}$.

\subsection{Spatial Emissivity Profile}
\label{subsect:SpatialEmissivity}

To determine the emission, 
we employ excitation rate coefficients calculated using the methods of Paper 1 (see also Appendix~\ref{sect:AppendixA}).  
An emission line photon is produced when a cold hydrogen atom or broad neutral is excited by a charged particle, 
or undergoes charge transfer to an 
excited state.  In the former case, the rates must be weighted by the probability of repeated excitation.  We use 
the following equations to calculate the spatial emissivity profiles for the narrow ($\xi_n$) and broad ($\xi_b$) 
components:
\begin{eqnarray}
\xi_n(z) & = & {n_H\over 1-P_{E_0}} \sum_{s=e}^{\alpha} n_s \tilde{R}_{\Delta n,E_0,s},\\
\xi_b(z) & = & n_H n_p \tilde{R}_{\Delta n,T_0^*,p} + \sum_{k=1}^N n_k\left(n_p \tilde{R}_{\Delta n,T^*,p}+
	{1\over 1-P_E}\sum_{s=e}^{\alpha} n_s \tilde{R}_{\Delta n,E,s}\right),
\end{eqnarray}
where the rate coefficients are labeled by the transition $\Delta n = H\alpha, Ly\alpha, Ly\beta$, atomic 
interaction, and particle type.  
The symbols $E_0$ and $T_0^*$ denote excitation and charge transfer to an excited state for 
cold hydrogen, while $E$ and $T^*$ 
denote these interactions for broad neutrals.  The probabilities $P_{E_0}=R_{E_0}/(R_{E_0}+R_{I_0}+R_{T_0})$ and 
$P_E=R_E/(R_E+R_I+R_T)$ are calculated using the total reaction rates per atom or broad neutral for ionization, 
excitation, and charge transfer.  These are calculated using the weighted sum of the rate coefficients, for example, 
$R_I = \sum_{k=1}^N n_k \sum_{s=e}^{\alpha} n_s \tilde{R}_{iB,s}$.

Figure~\ref{fig:emb1f1v1000} shows the spatial emissivity profiles 
for the narrow and broad 
components as a function of distance $z$ behind the shock front.
The thin vertical lines indicate the centroids 
$z_{cn,b}$ of the emission components, calculated according to the formula:
\begin{eqnarray}
\label{eq:centroid}
\int_0^{z_{cn,b}} dz\, \xi_{n,b}(z) - \int_{z_{cn,b}}^{\infty} dz\, \xi_{n,b}(z) = 0.
\end{eqnarray}
For the low value of $f_p$ shown in this figure, significant ionization must occur before there are enough 
charged particles to excite the narrow emission and engage in charge transfers to produce broad neutrals.  
Therefore, the intensity peaks for both components are shifted downstream from the shock front, and the 
centroid of the broad line emission is shifted farther downstream than that of the narrow line.

\subsection{Line Profiles}
\label{subsect:LineVelocity}

The full width half maximum (FWHM) of the broad line can be related to the velocity and temperature equilibration 
of the shock.  
The line profile is a convolution of 
the broad neutral and exciting species distribution functions with the cross sections for excitation and 
charge transfer to an excited state, projected along the line of sight to the observer.
In a cylindrical coordinate 
system $(r,\theta,z)$ where the $z$ axis is along the shock velocity direction, it is straightforward to project 
the line profile for observers oriented both edge-wise (along the $r$ axis) and face-on (along the $z$ axis) with respect 
to the shock front.  Due to limb brightening, most observations will be selected for shocks viewed edge-wise or nearly so.

We calculate hydrogen line profiles $\phi_{FO}$, and $\phi_{EW}$ according 
to the formulas:
\begin{eqnarray}
\label{eq:phiFO}
\phi_{FO}(v_z) & = & n_H n_p F_{Tp,z}(v_s) + \nonumber\\
	& & \int d^3 {\bf v}' \left[
	\sum_{s=e}^{\alpha} n_s f_s({\bf v}') F_{Es,z}(v_z,{\bf v}')+n_p f_p({\bf v}') F_{Tp,z}(v_z,{\bf v}')\right],\\
\label{eq:phiEW}
\phi_{EW}(v_r,z) & = & n_H n_p F_{Tp,r}(v_s) + \nonumber\\
	& & \int d^3 {\bf v}' \left[
	\sum_{s=e}^{\alpha} n_s f_s({\bf v}') F_{Es,r}(v_r,{\bf v}') + n_p f_p({\bf v}') F_{Tp,r}(v_r,{\bf v}')\right],\\
F_{Xs,z} & = & \int_0^{2\pi} d\theta \int_0^{\infty} d v_r v_r \sum_{k=1}^N n_k\, f_k({\bf v})\, \Delta v\, 
	\sigma_{X,H\alpha,s}(\Delta v),\\
\label{eq:phiFr}
F_{Xs,r} & = & v_r \int_0^{2\pi} d\theta \int_{-\infty}^{\infty} dv_z \sum_{k=1}^N n_k\, f_k({\bf v})\, \Delta v\, 
	\sigma_{X,H\alpha,s}(\Delta v),
\end{eqnarray}
where $\Delta v\equiv |{\bf v}-{\bf v}'|$, and the cross sections $\sigma_{X,H\alpha,s}$ denote excitation and 
charge transfer reactions for H$\alpha$ emission to appropriately weighted angular momentum states, using 
eqs.~(\ref{eq:sigHalpha}) and (\ref{eq:fitfrac}).  It should be noted that in eqs.~(\ref{eq:phiFO})--(\ref{eq:phiFr}), 
the quantities denoted by $f$ are kinetic distribution functions, and should not be confused with the pre-shock 
ionization fraction $f_p$.
The edge-wise profile can be calculated as a function of position behind the shock.  
In practice, however, the observed lines are not spatially 
resolved.  We therefore calculate the profiles as a function of $v_s$ and $\beta$, spatially averaged over the 
shock transition zone.  
Figure~\ref{fig:profiles} shows examples of symmetric, edge-wise broad neutral velocity distributions, which are 
inputs to eq.~(\ref{eq:phiEW}).
Results are depicted in a reference frame where the average ion velocity is zero, 
for $\beta=0.01,0.1,0.5,1$, with $v_s=1000$ km s$^{-1}$ (left panel), 
and $7000$ km s$^{-1}$ (right panel).  For all cases, the pre-shock 
ionization fraction is set to $f_p=0.5$.

For $v_s\la 1500$ km s$^{-1}$, excitation by electrons dominates over that by protons 
(including charge transfer into excited states).  Even at 
moderate values of the electron temperature, $\beta\la 0.1$, the thermal width of the electron distribution function is 
much larger than that of the broad neutral distribution, and the electrons typically have much higher velocities.  Thus, 
for the majority of the range of integration, $\Delta v\approx v'$, and the electron contribution to 
eqs.~(\ref{eq:phiFO})--(\ref{eq:phiEW}) is approximately equal to the projected velocity distribution of the broad 
neutrals, multiplied by the rate coefficient for excitation by electrons.  This approximation has been used in 
previous studies of the line profile \citep[e.g.,][]{Chevalieretal80a,Ghavamianetal01a,HengMcCray07a}.  Nevertheless, for 
shock velocities $v_s\sim 2000$ km s$^{-1}$, excitation rates by electrons and protons are comparable, with the 
proton contribution increasing for faster shocks.  The cross sections for excitation by protons 
increase with the relative speed of the colliding particles.  Since the broad neutrals and ions have comparable 
speeds, high speed neutrals are more likely to produce H$\alpha$ photons.  The effect on the line profile is 
somewhat mitigated by the integration over velocity space, but the observed line width is larger than the 
velocity width of the broad neutral distribution.  This effect is relatively small at low velocity ($<10\%$) but is 
significant for high velocity shocks (see \S\ref{sect:Discussion}).

Figure~\ref{fig:FWHM} displays the FWHM of the broad, edge-wise H$\alpha$ line profile 
as a function of $v_s$, 
at several values of $\beta=0.01,0.1,0.5,1$.  The pre-shock ionization is set at $f_p=0.5$.
The FWHM increases monotonically with $v_s$ due to the increased temperature of the post-shock 
ion distributions (and hence broad neutral distributions) behind faster moving shocks.  
As $\beta$ is increased, energy is transferred from the protons to 
heat the electrons, leading to lower proton temperatures and a smaller FWHM for the broad component.

For fast shocks, in which proton excitation contributes substantially to eqs.~(\ref{eq:phiFO})--(\ref{eq:phiEW}), 
we expect different transitions 
(e.g., H$\alpha$ and Ly$\alpha$) to have different FWHM relations, since they employ distinct reaction cross sections 
in the calculation of $\phi$.  This is a unique prediction made by our calculation, and 
may have consequences for studies of Ly$\alpha$ emission from shocks with $v_s\ga 4000$ km s$^{-1}$, in which  
the FWHM of the Ly$\alpha$ line increases from 10-60\% over that of the H$\alpha$ line.

\section{Dependence on Shock Parameters}
\label{sect:Results}

Here we describe how the structure and emission from the transition zone of Balmer-dominated SNRs depend 
on shock velocities (ranging from $v_s=300-10,000$ km s$^{-1}$), pre-shock ionization fractions 
(ranging from $f_p=0.1-0.9$), and 
temperature equilibration ratios (ranging from $\beta=0.1-1$), as well as Case A and Case B conditions for the 
narrow line.

\subsection{Spatial Structure of the Shock Transition Zone}
\label{subsect:SpatialStructureResults}

The basic features of the shock transition zone density structure were shown in Figure~\ref{fig:denb1f5v1000}.  The 
details of broad neutral production are highly sensitive to $v_s$, $f_p$, and $\beta$.  
We can see the effect of increasing shock velocity by comparing Figure~\ref{fig:denb1f5v4000} to 
Figure~\ref{fig:denb1f5v1000}.
At high $v_s$, the ionization rate is greater than the 
charge transfer rate, so that relatively few broad neutrals are produced.  
The densities of the broad neutral populations decrease rapidly with each subsequent 
charge transfer reaction.  The helium ionization rates decrease, causing the 
$He^+$ and alpha particle production to peak further downstream.

In Paper 2, we explored variations in ion velocity throughout the shock transition zone and found that, for 
$v_s\ga 300$ km/s, $v_i\approx v_s/4$ with negligible deviation.  We confirm this conclusion with our multi-component 
models.
Figure~\ref{fig:vsb10f05} shows the dimensionless ion velocity and temperature as a function of distance from the 
shock front.  Results are shown for $v_s=1000$ km s$^{-1}$ and $v_s=4000$ km s$^{-1}$, with $f_p=0.5$, $\beta=1$.  
In the left panel, we 
display the percent deviation of the ion velocity from $v_s/4$ for the two shock velocities.  At $v_s<1600$~km~s$^{-1}$, 
broad neutrals are produced with smaller 
momentum density than the original ion population, leading to an increase in ion velocity by conservation of momentum 
(solid curve).  For $v_s\ga 1600$ km~s$^{-1}$, the opposite occurs, and the ion velocity decreases as the broads are 
produced (dotted curve).  The deviation of ion velocity from $v_s/4$ is less than $1\%$ and has a negligible effect 
on calculations of the reaction rate coefficients .  
In the right panel, we plot the ion temperature profile.  In this case, the broad neutrals 
are produced 
with slightly lower temperature than the ions, leading to slight heating of the ions.  At large distances 
downstream, the presence of alpha particles increases the mean atomic mass in the gas to 1.27 $m_p$ (assuming a 
$10\%$ helium abundance), leading to a final ion temperature 
that is slightly higher than $\epsilon_i\sim 3/16(1+\beta)$.  The maximum deviation of $\epsilon_i$ from its 
expected value is of order $10\%$, which in practice has a negligible effect on the values of the reaction rate 
coefficients.

\subsection{H$\alpha$ Emissivity}
\label{subsect:EmissivityResults}

\begin{figure}
\begin{center}
\plotone{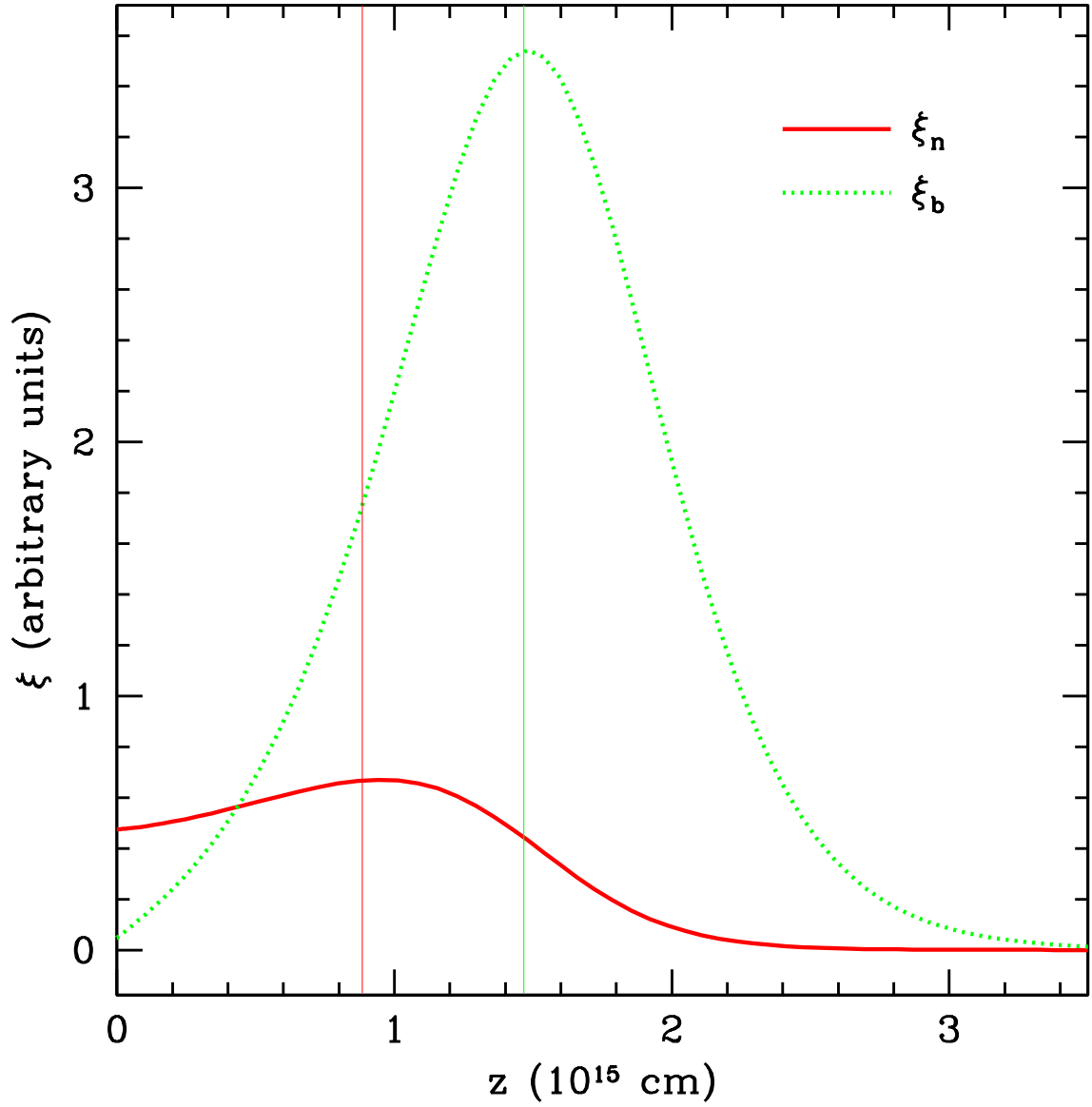}
\end{center}
\caption{Emissivity profiles for narrow and broad emission as a function of 
distance behind the shock front.  Results are shown for $v_s=1000$ km~s$^{-1}$, 
$f_p=0.1$, $\beta=1$, and $f_{\rm He}=0.1$. 
The centroids of the narrow and broad profiles are 
indicated by the light vertical lines.}
\label{fig:emb1f1v1000}
\end{figure}

\begin{figure}
\begin{center}
\plotone{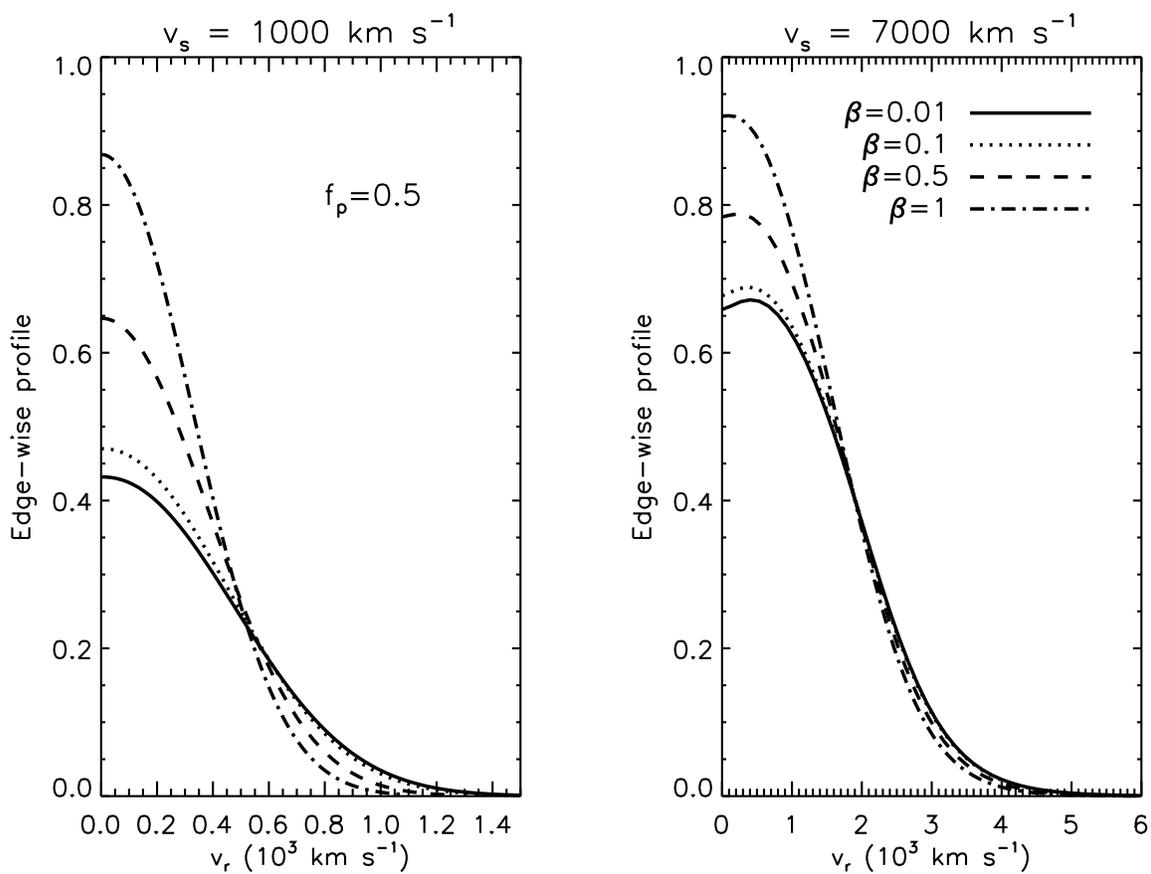}
\end{center}
\caption{Symmetric, edge-wise broad neutral velocity distributions as a function of line-of-sight velocity $v_r$, 
in a reference frame 
where the proton velocity is zero.  Results are shown 
for fixed $f_p=0.5$ and several values of $\beta=0.01,0.1,0.5,1$.  As $\beta$ is increased, the temperatures of the proton 
and broad neutral distribution functions decrease, leading to smaller FWHM.}
\label{fig:profiles}
\end{figure}

\begin{figure}
\begin{center}
\plotone{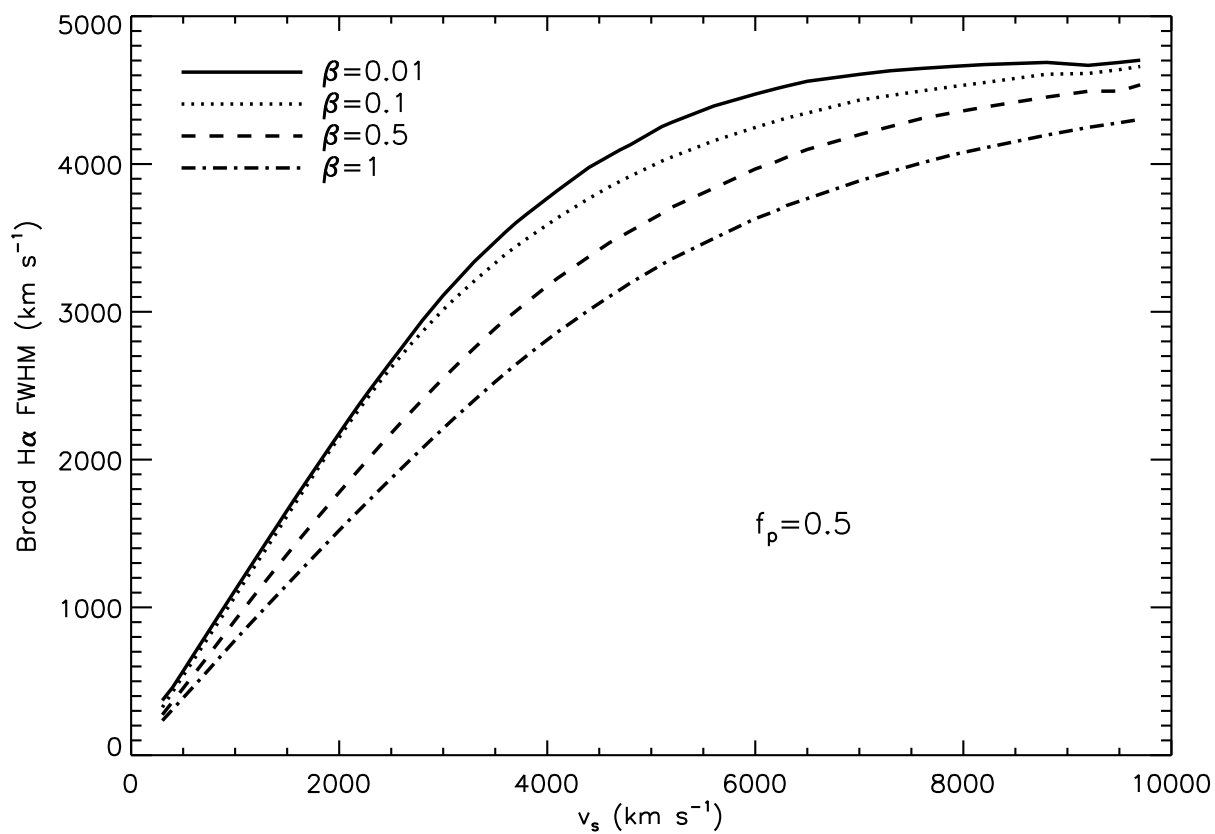}
\end{center}
\caption{Broad, edge-wise H$\alpha$ line FWHM as a function of shock velocity for fixed $f_p=0.5$ at several values of 
$\beta=0.01,0.1,0.5,1$ for H$\alpha$ transitions.  
As $\beta$ is increased, the predicted FWHM decreases (see Figure~\ref{fig:profiles}).  
}
\label{fig:FWHM}
\end{figure}

\begin{figure}
\begin{center}
\plottwo{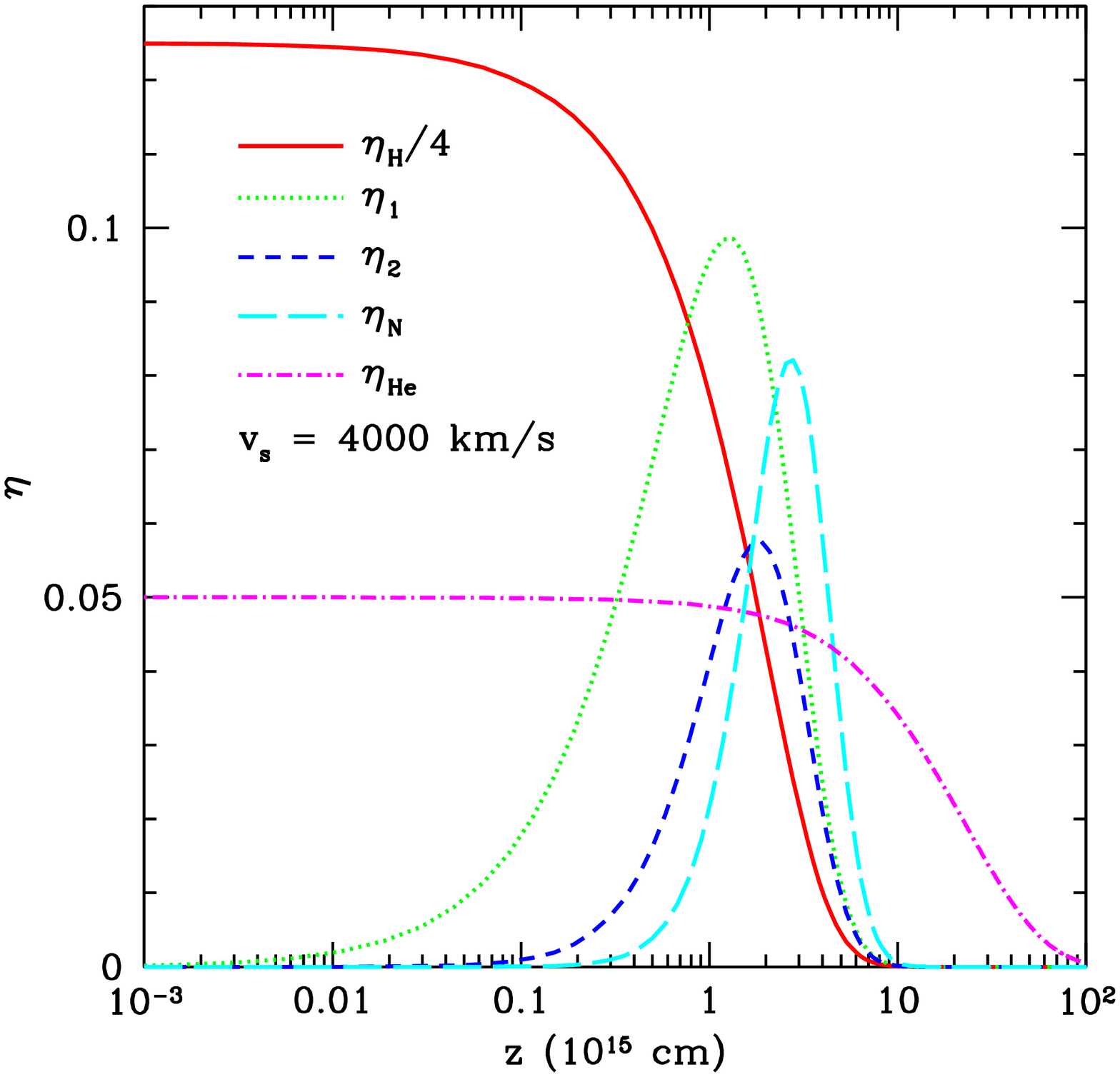}{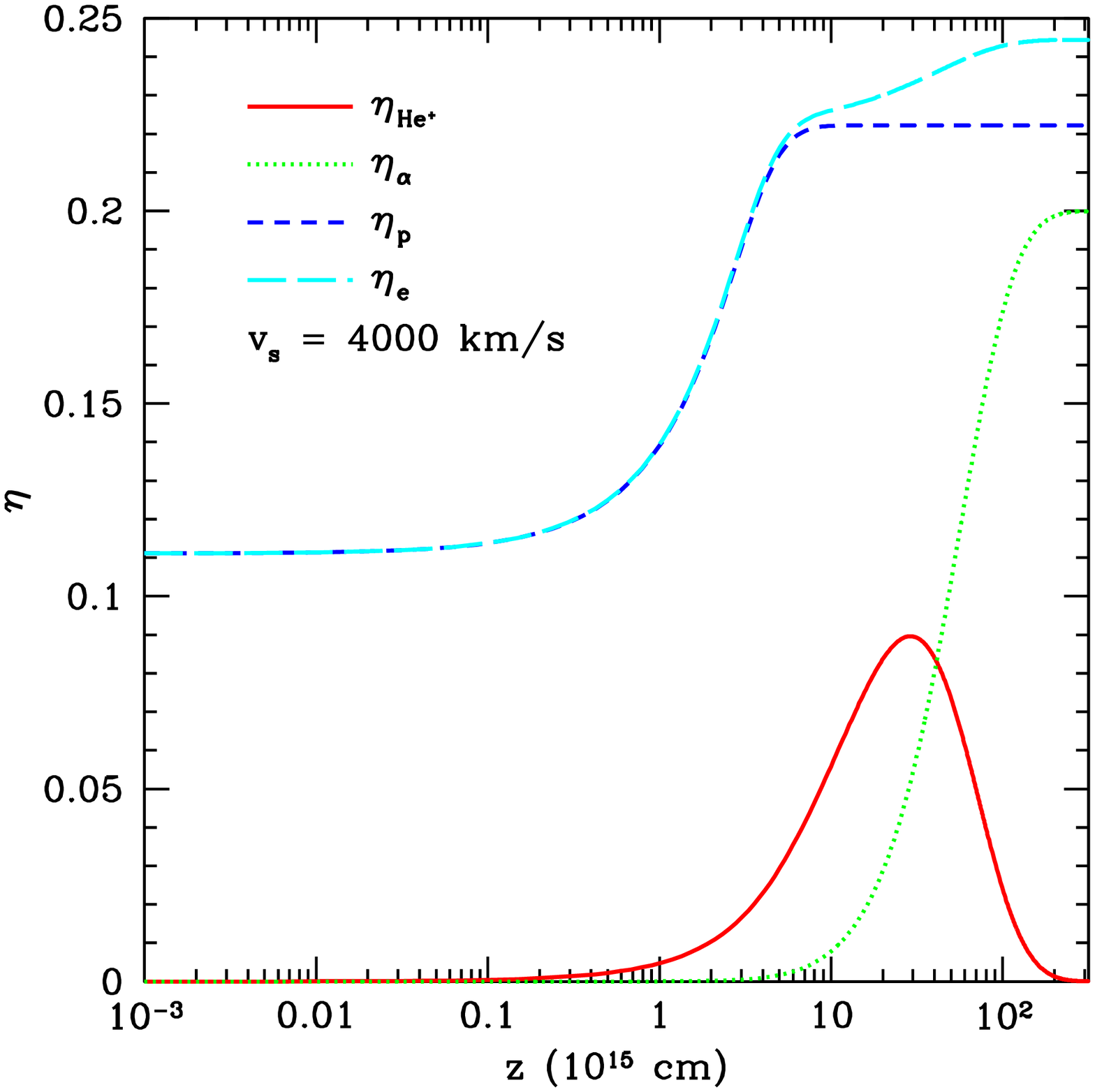}
\end{center}
\caption{Same as Figure~\ref{fig:denb1f5v1000} except that $v_s=4000$ km s$^{-1}$. 
The density of the cold neutral hydrogen is scaled by $1/4$.}
\label{fig:denb1f5v4000}
\end{figure}

\begin{figure}
\begin{center}
\plottwo{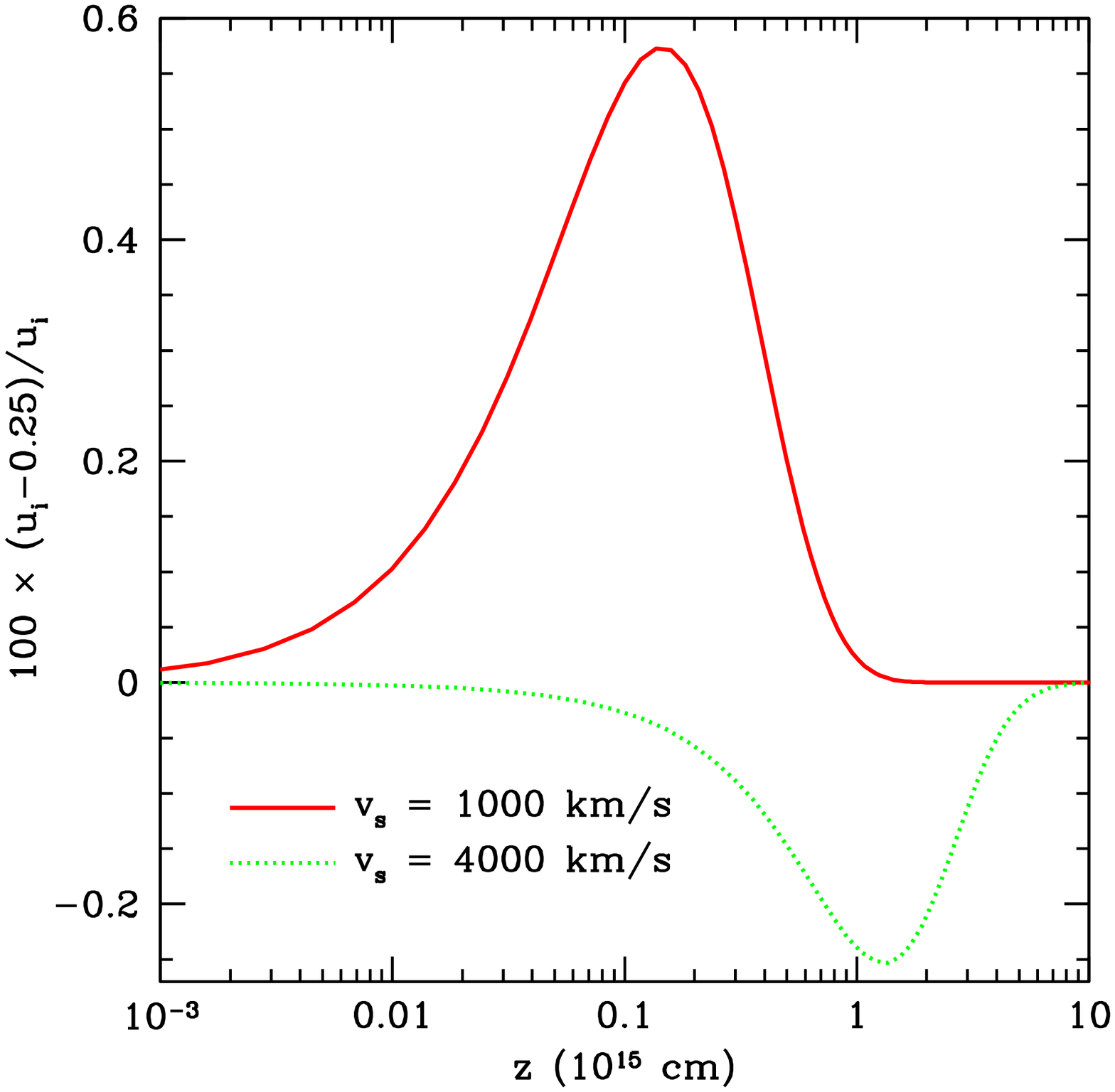}{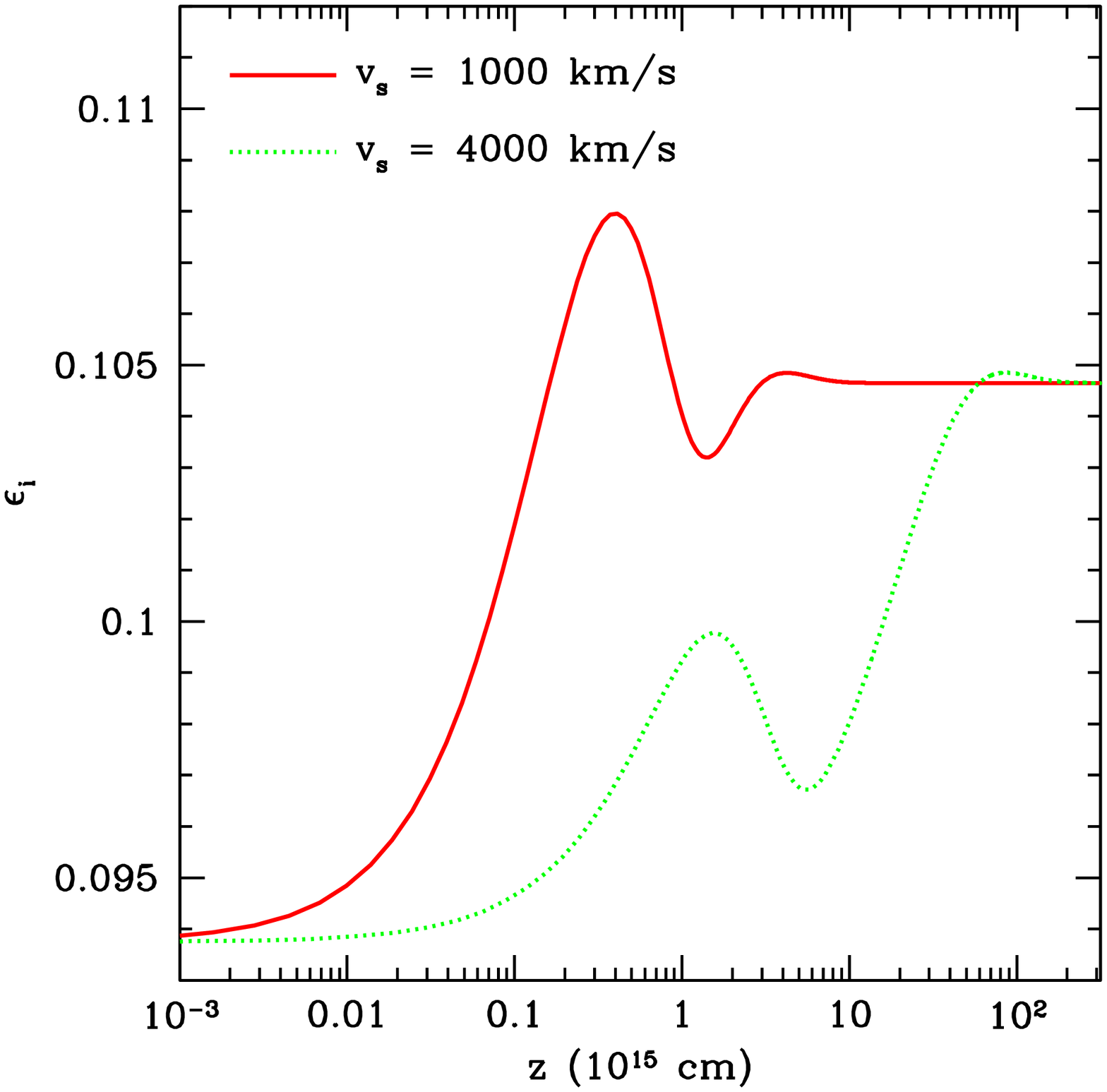}
\end{center}
\caption{Percent deviation of the ion velocity from $v_s/4$ (left panel) and 
the ion temperature (right panel) as a function of position behind the 
shock front for two values of the shock velocity: $v_s=1000$ km s$^{-1}$ 
(solid curve) and $v_s=4000$ km s$^{-1}$ (dotted curve).  Results are shown for 
$f_p=0.5$, $\beta=1$, and $f_{\rm He}=0.1$.}
\label{fig:vsb10f05}
\end{figure}

\begin{figure}
\begin{center}
\plottwo{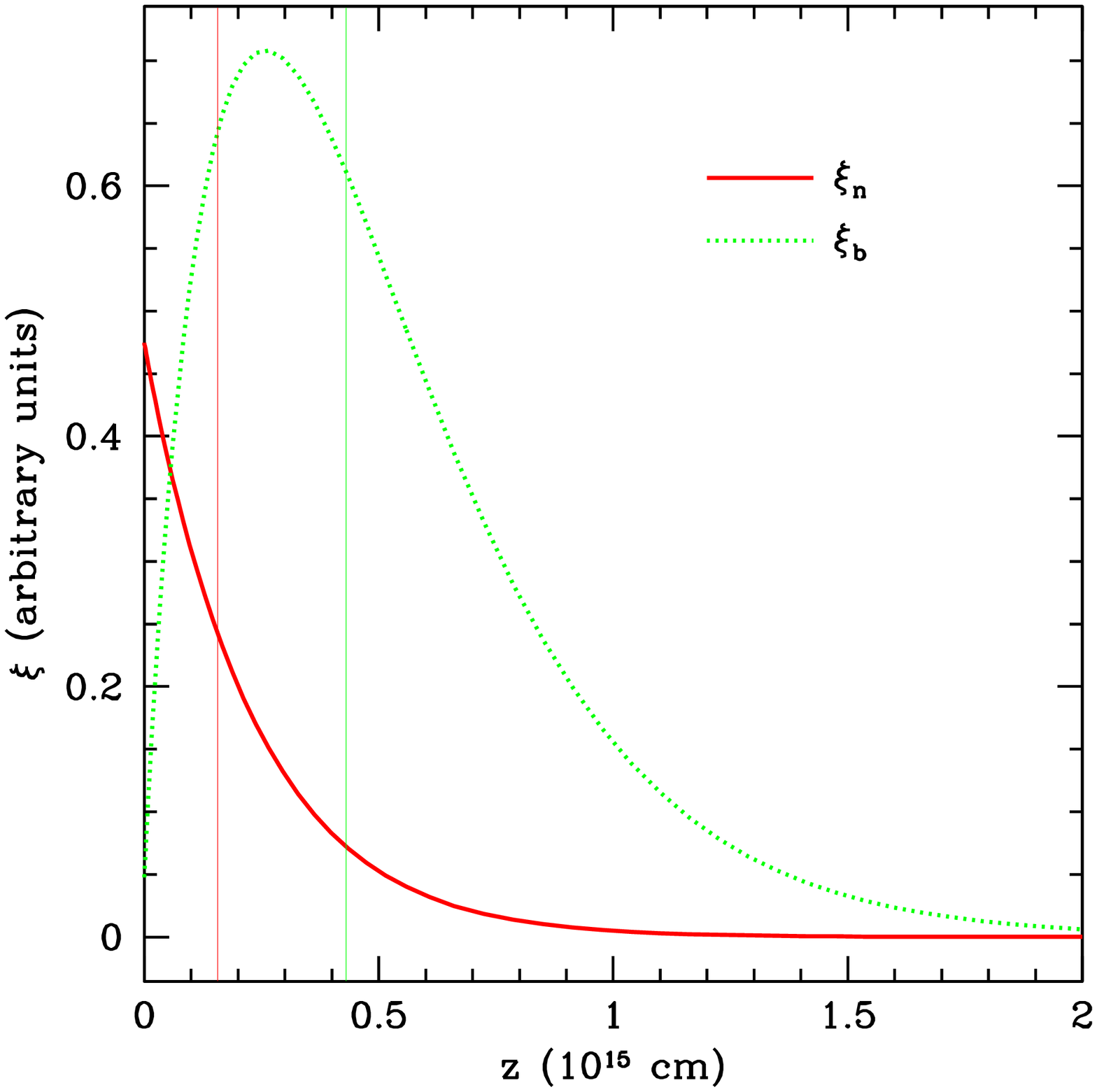}{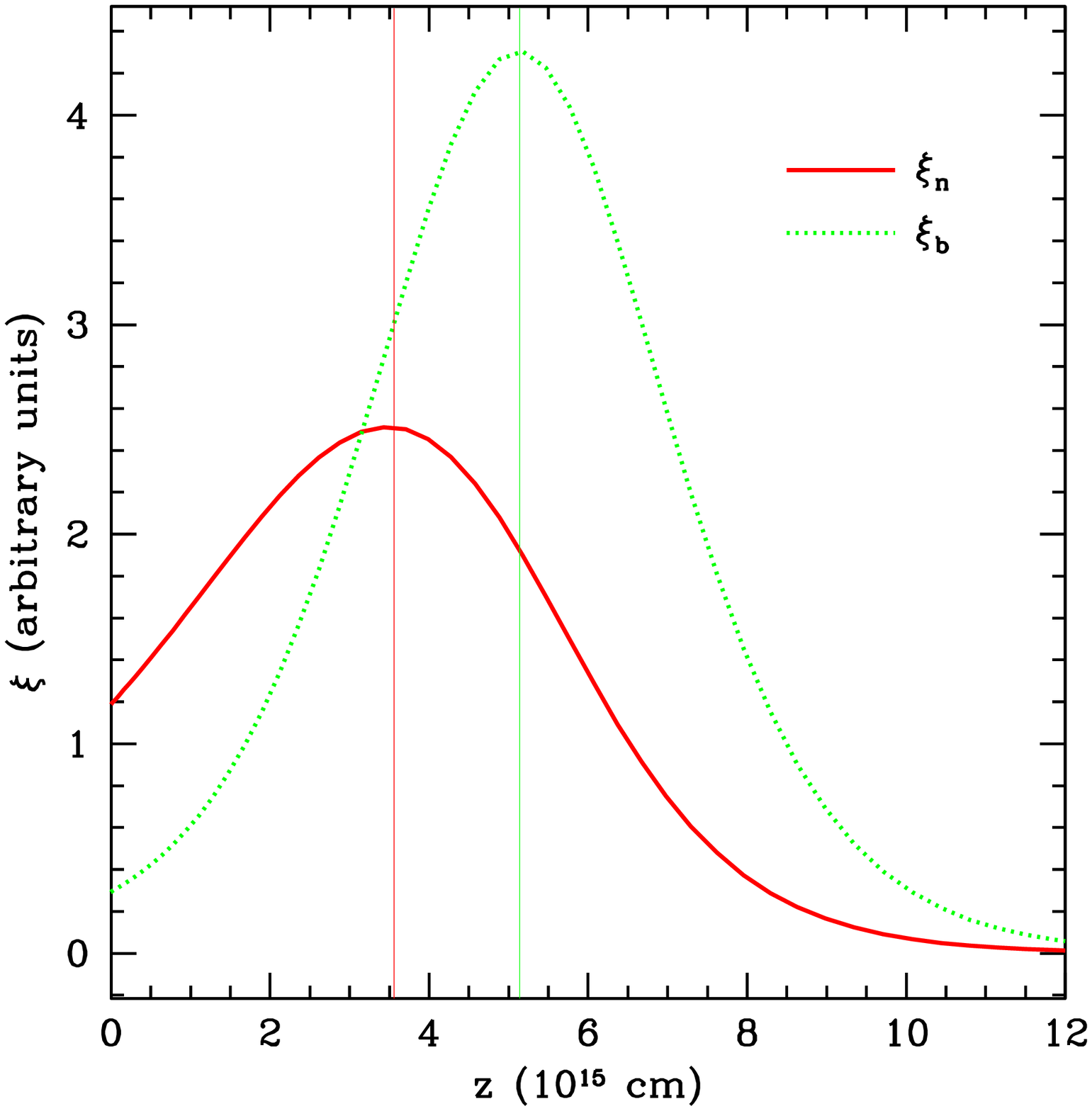}
\end{center}
\caption{Same as Figure~\ref{fig:emb1f1v1000}, except that $f_p=0.9$, $v_s=1000$ km s$^{-1}$ (left panel) and 
$f_p=0.1$, $v_s=4000$ km s$^{-1}$ (right panel).}
\label{fig:emv1b10f09}
\end{figure}

In Figure~\ref{fig:emb1f1v1000}, we show a typical emissivity profile at relatively low shock velocity and ionization 
fraction.  In the left panel of Figure~\ref{fig:emv1b10f09}, we show the effect of increasing the initial ionization 
fraction to $f_p=0.9$.  We see that  
the neutral density decreases rapidly and the narrow line emission peaks at the shock front (left panel).  
Many charge transfer reactions occur close to the shock, pushing the centroid of the broad emission farther 
upstream compared to the 
$f_p=0.1$ case.  Comparing the right panel of Figure~\ref{fig:emv1b10f09} to Figure~\ref{fig:emb1f1v1000}, we can see the 
effect of increasing the shock velocity.  For the high velocity case, both ionization and charge transfer rates are 
decreased, shifting the centroids of both line components farther downstream in the transition zone.  

\begin{figure}
\begin{center}
\plottwo{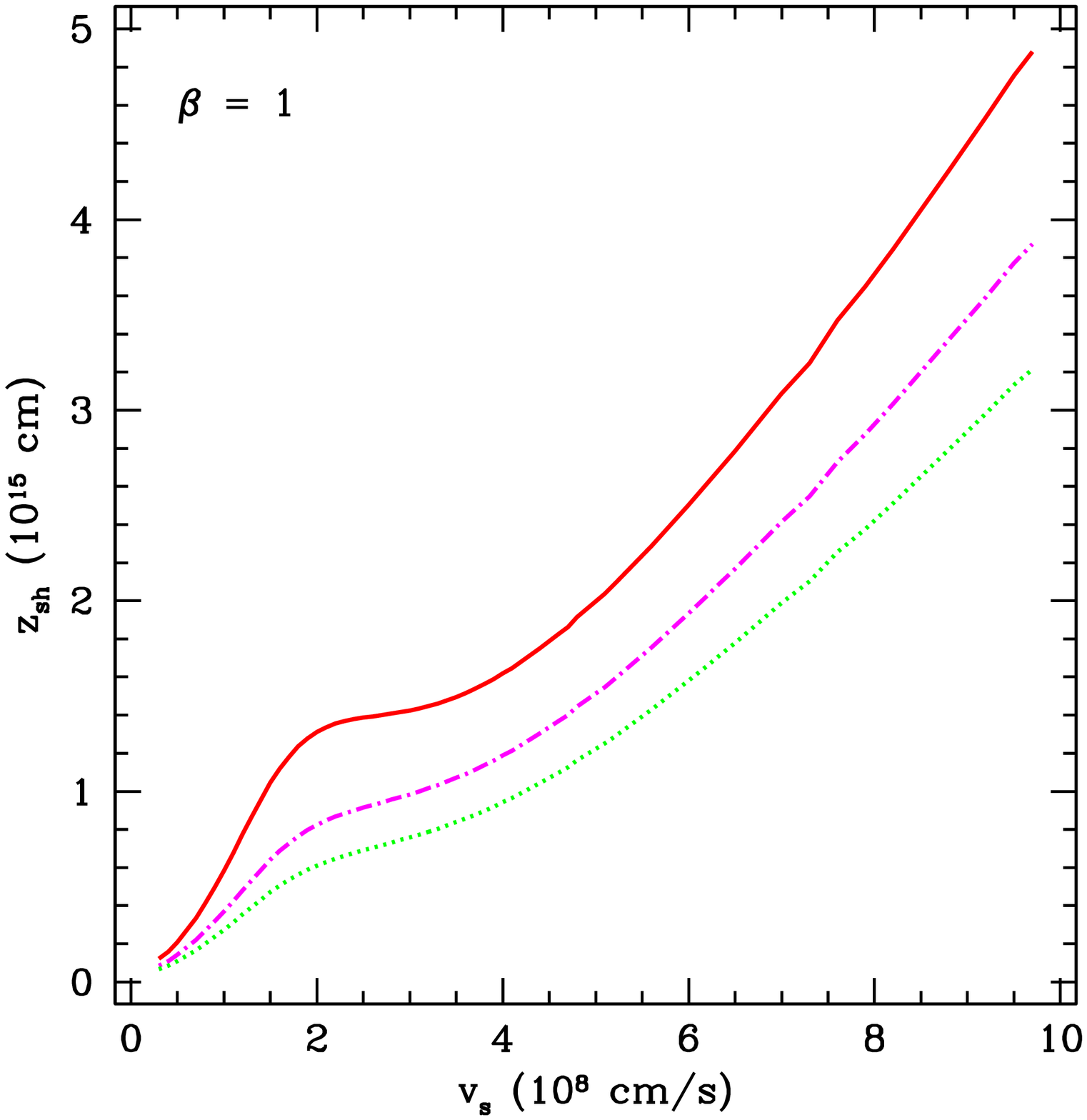}{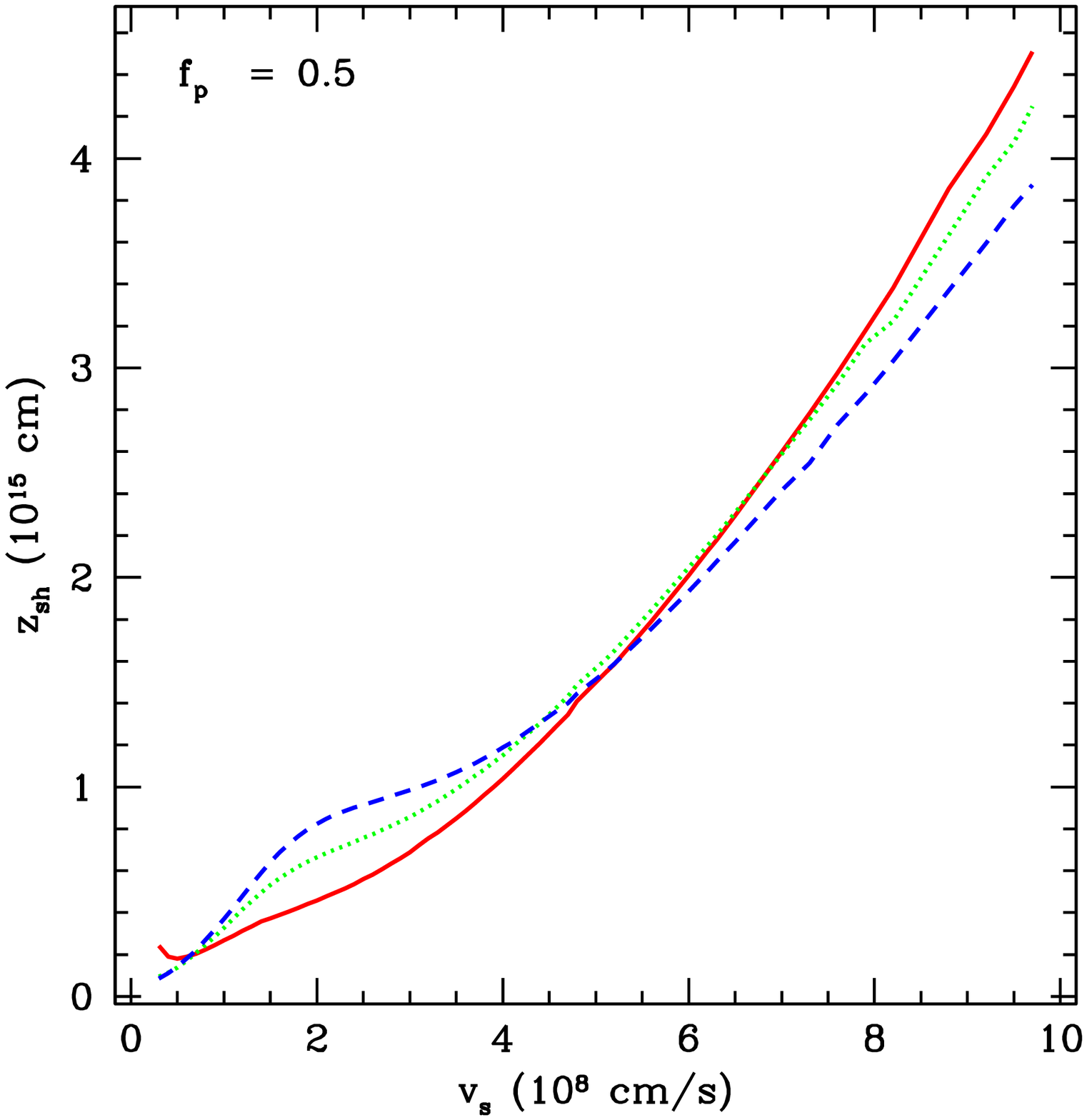}
\end{center}
\caption{Spatial shift in units of $10^{15}$ cm as a function of shock 
velocity $v_s$.  The pre-shock density is set to $n_0=1$ cm$^{-3}$; note that 
$z_{sh}$ scales as $1/n_0$.  The left panel shows 
the variation of the shift with $f_p$ at fixed $\beta=1$.
From top to bottom, the curves show results 
for pre-shock ionization fractions $f_p=0.1, 0.5, 0.9$.
The right panel shows the variation of the shift with $\beta$, for fixed 
$f_p=0.5$.  From bottom  
to top at $v_s\sim 2000$ km s$^{-1}$, the curves show results for temperature 
equilibration ratios $\beta=0.1,0.5,1$.}
\label{fig:shift}
\end{figure}

As discussed by \citet[][]{Raymondetal07a} and Paper 2, the spatial shift between the broad and narrow line centroids 
potentially provides a 
constraint on the pre-shock ionization fraction, $f_p$, and external density, $n_0$.  In Figure~\ref{fig:shift}, we plot 
$z_{\rm sh}\equiv z_{cb}-z_{cn}$ as a function of shock velocity $v_s$, with $n_0=1$ cm$^{-3}$.  Note that 
the spatial shift scales as $z\propto 1/n_0$.  
In the left panel, we show how $z_{\rm sh}$ depends on 
$f_p$, holding $\beta=1$.  
As the velocity increases, the charge transfer rate decreases 
relative to the ionization rate, delaying the production of broad neutrals.  Consequently, the centroid of 
$\xi_b$ shifts downstream and the value of $z_{\rm sh}$ increases.  For small $f_p$, few protons initially exist to 
engage in charge exchange.  Consequently,  $z_{cb}$ shifts downstream from the shock front.  In the right panel, 
we show how the results depend on the temperature equilibration parameter, $\beta$, for fixed $f_p=0.5$.  
At shock velocities $v_s\ga 1000$ km s$^{-1}$, increasing $\beta$ reduces the 
proton ionization rate for cold hydrogen. 
This effect causes the peak of broad production to shift downstream from shock front.  
At high velocities, increasing 
$\beta$ tends to increase the charge transfer rate for broad neutrals over that for cold hydrogen. 
This effect causes the peak of broad production to shift closer to the shock front, and decreases the value 
of $z_{\rm sh}$.  
For more discussion on the 
observational significance of the spatial emissivity profile and the shift, see \S\ref{subsect:Limitations}.

\begin{figure}
\begin{center}
\plottwo{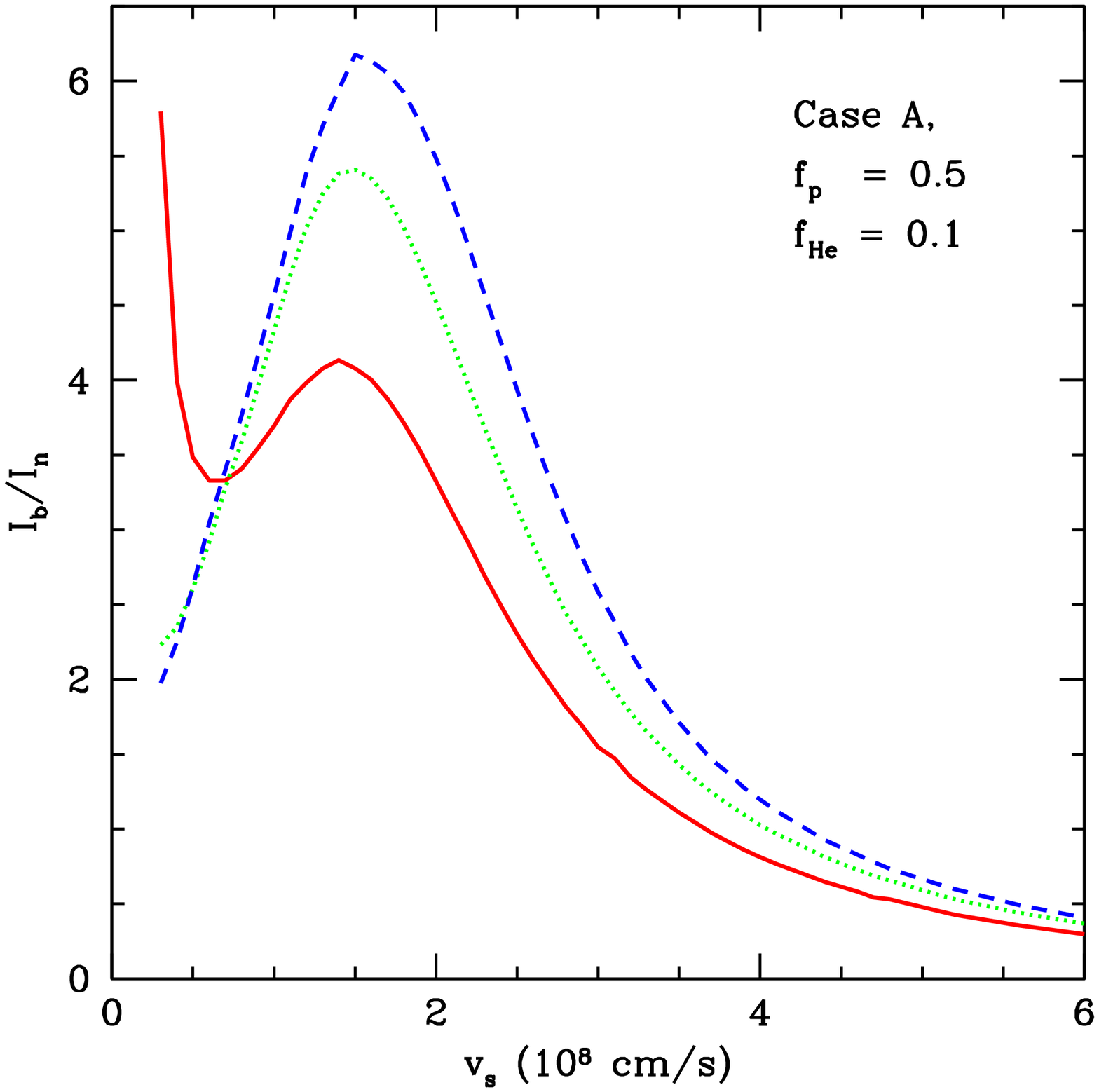}{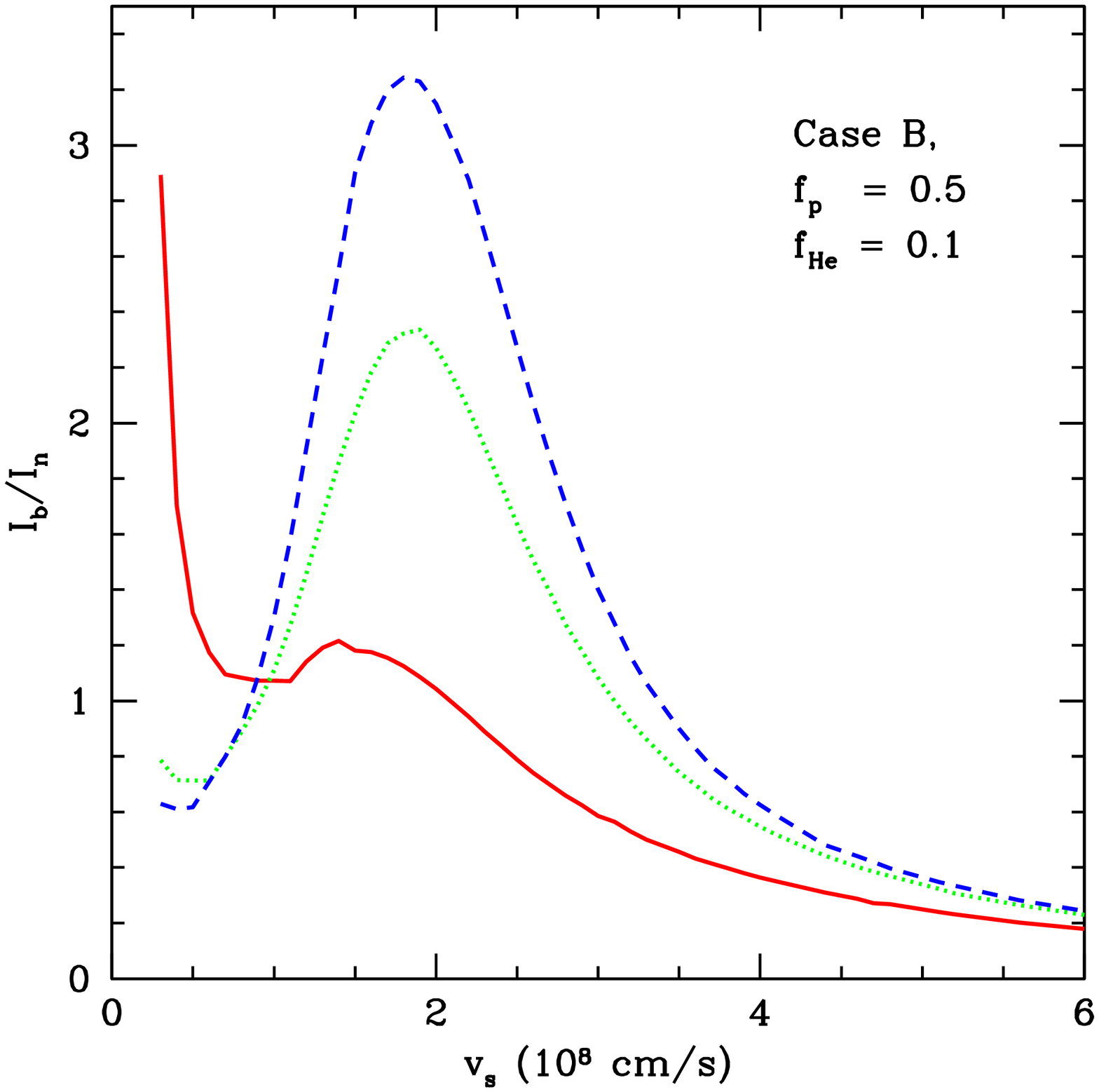}
\end{center}
\caption{Broad to narrow intensity ratio $I_b/I_n$ 
as a function of shock velocity 
$v_s$ with $f_p=0.5$, $f_{\rm He}=0.1$ for Case A (left panel) and 
Case B (right panel) conditions.  
From bottom to top, the 
curves show results for temperature equilibration ratios $\beta=0.1,0.5,1$.}
\label{fig:ibin}
\end{figure}

Another important observational diagnostic for Balmer-dominated SNRs is the ratio of 
integrated broad-to-narrow line strengths, defined as: 
\begin{eqnarray}
\label{eq:ibin}
I_b/I_n = {\int_0^{\infty} dz\, \xi_b(z)\over \int_0^{\infty} dz\, \xi_n(z)}.
\end{eqnarray}
This ratio has a strong dependence on both $v_s$ and $\beta$.  Figure~\ref{fig:ibin} shows $I_b/I_n$ versus 
shock velocity at fixed $f_p=0.5$, for several values of temperature equilibration $\beta=0.1,0.5,1$, using 
Case A (left panel) and Case B (right panel) conditions for the narrow line.  For a 
given $\beta$, the variation of the intensity ratio with velocity is the result of competition between 
charge transfer and ionization, which contribute equally at $v_s\sim 2000$ km s$^{-1}$.  At very low $v_s$, the 
ionization rate begins to decrease precipitously while the charge transfer rate stays roughly constant, leading to the 
spike in $I_b/I_n$.  At a fixed shock velocity, increasing $\beta$ causes a decrease in the proton ionization rate 
for broad neutrals compared to cold hydrogen, leading to an increase in the intensity ratio.  
In Case B conditions (right), 
additional narrow emission due to absorption of trapped Ly$\beta$ photons decreases the values of 
$I_b/I_n$ relative to Case A.  As noted in 
\S\ref{sect:EmissionPhysics}, for most Balmer-dominated SNRs, partial line scattering of Ly$\beta$ photons yields 
emission intermediate between Cases A and B.

\begin{figure}
\begin{center}
\plottwo{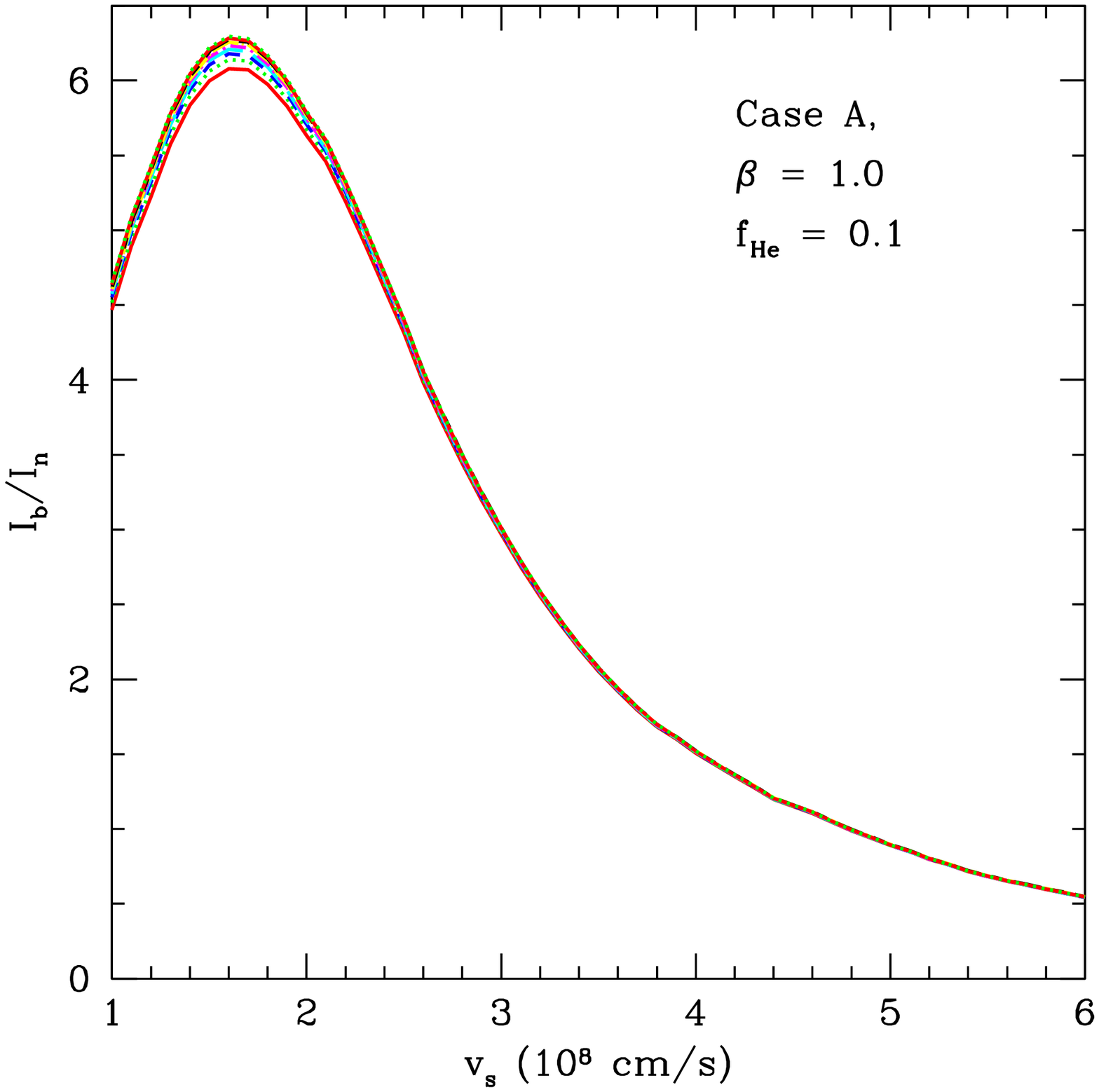}{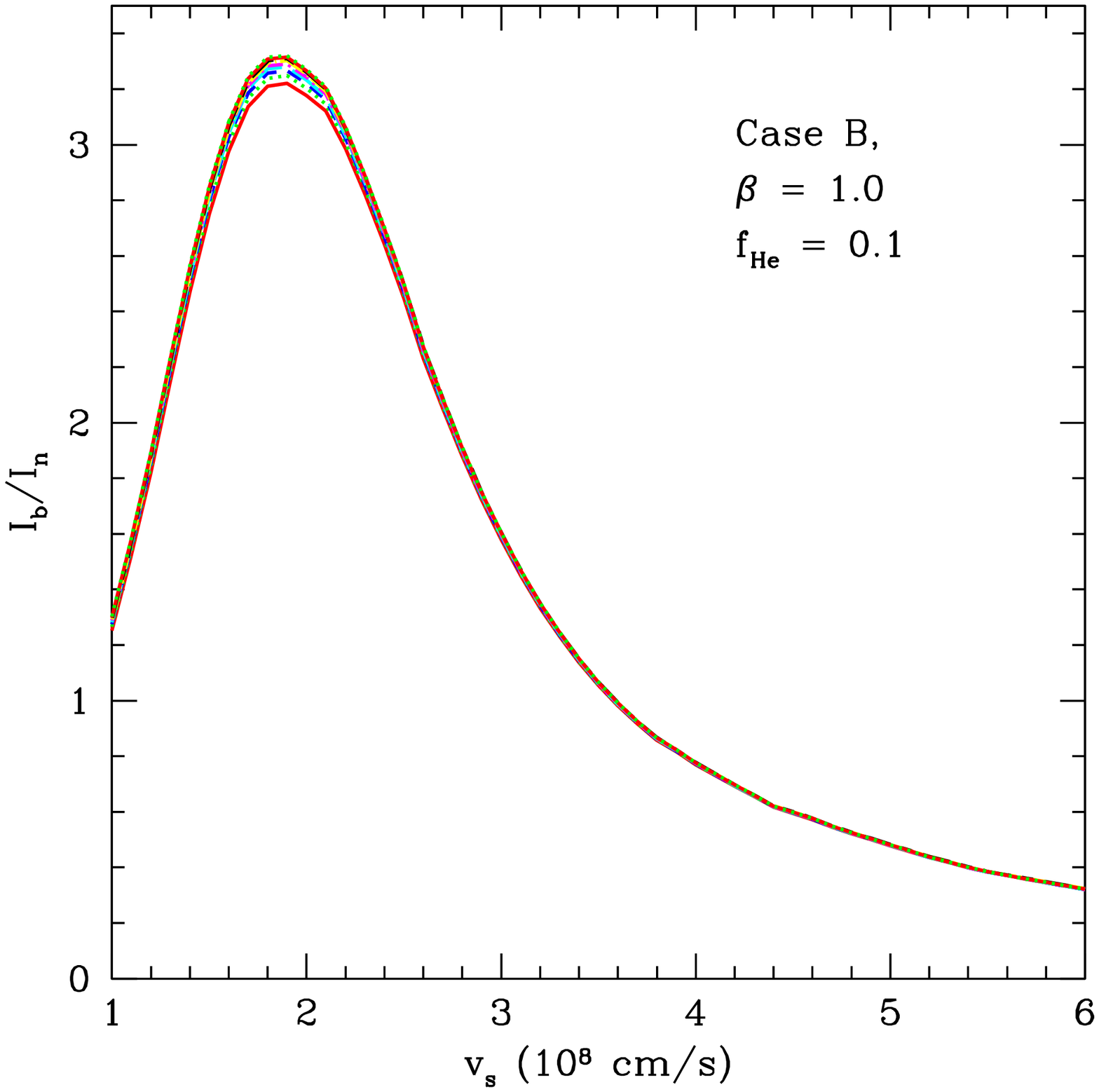}
\end{center}
\caption{Broad to narrow intensity ratio $I_b/I_n$ 
as a function of shock velocity 
$v_s$, with $\beta=1$, $f_{\rm He}=0.1$, Case A (left panel) and 
Case B (right panel) conditions.  
From bottom to top, the 
curves show results for pre-shock ionization fraction $f_p=0.1,0.5,0.9$.}
\label{fig:ibinfp}
\end{figure}

According to the picture in Paper 1,
the dependence of $I_b/I_n$ on $f_p$ is expected to be weak, since the number of broad 
neutrals produced in each population per cold hydrogen atom is fixed.  This intuition is confirmed by our 
multi-component calculation.  When $f_p$ is increased, we expect more charge transfers in the shock transition zone.  
However, this effect is balanced by higher ionization rates for both broad and cold neutral populations.  
The net result is a negligible change in the $I_b/I_n$ ratio.  The presence of 
neutral helium introduces a weak dependence of $I_b/I_n$ on $f_p$.
In Figure~\ref{fig:ibinfp}, we show $I_b/I_n$ as a function of $v_s$ for fixed $\beta=1$ in Case A (left panel) and 
Case B (right panel) conditions.  
As $f_p$ decreases, fewer protons 
engage in charge transfer and ionization reactions close to the shock front, shifting the peak of broad neutral 
production downstream.  In this case, the broad neutrals persist far enough downstream to interact with the charged 
helium species produced there, altering the ratio $I_b/I_n$. This effect is of order $\la 1\%$ for 
$f_{He}=0.1$ and remains small for larger helium fractions.  However, it should be noted that, when the effects of 
partial Ly$\beta$ scattering are included, variations in optical depth with $f_p$ can introduce a 
dependence of $I_b/I_n$ on pre-shock ionization fraction \citep[][]{Ghavamianetal01a}.  While we have roughly treated 
Ly$\beta$ scattering using the prescription of \S\ref{sect:EmissionPhysics}, a full calculation of line scattering in 
the shock transition zone is needed to properly take into account these effects and incorporate the dependence of the 
integrated line ratio on the pre-shock ionization fraction.  This is a source of systematic error in 
our calculation.

Neglecting the $f_p$ dependence, the broad-to-narrow intensity ratio can be written as a function
\begin{eqnarray}
\label{eq:ibintheory}
I_b/I_n = L_{bn}(v_s,\beta),
\end{eqnarray}
where $I_b/I_n$ and $L_{bn}$ represent the measured and theoretical values of the intensity ratio, respectively.

\subsection{Broad Line Profiles and Observational Interpretations}
\label{subsect:BroadLineResults}

\begin{figure}
\begin{center}
\plotone{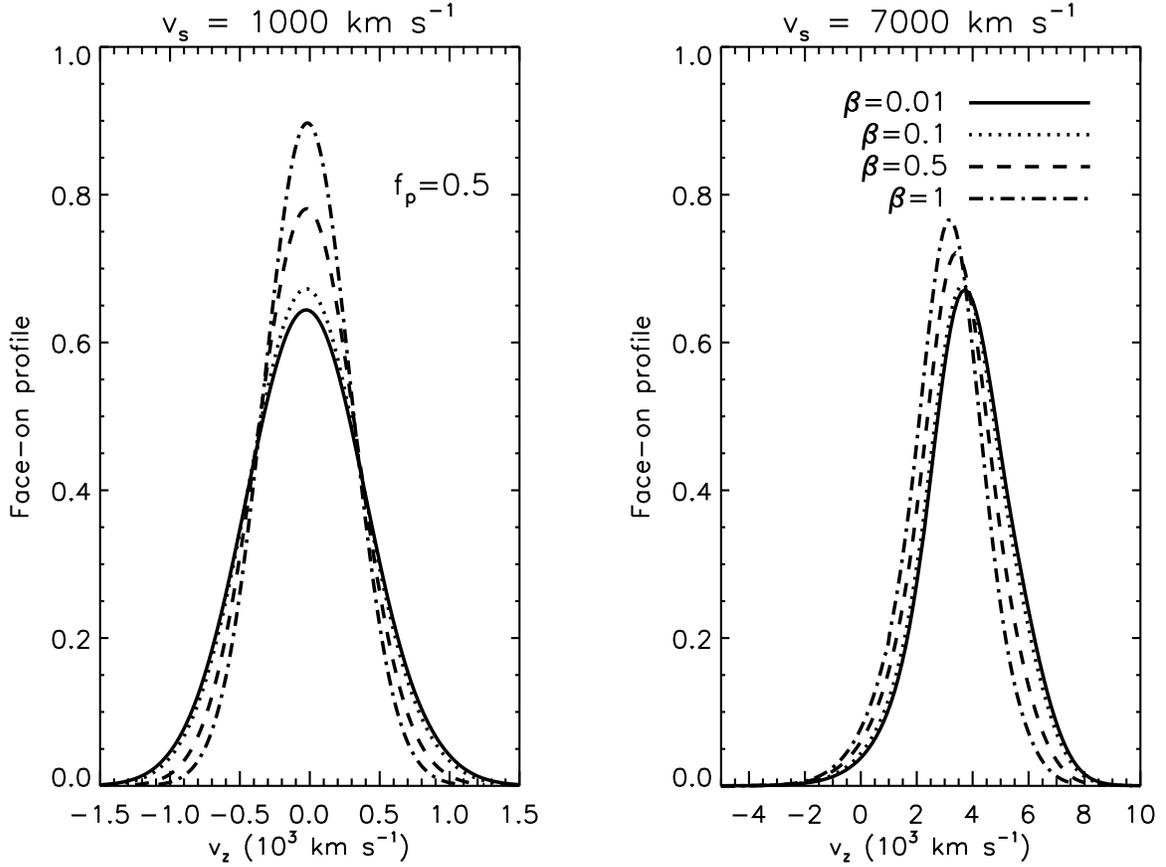}
\end{center}
\caption{Face-on broad neutral velocity distributions as a function of line-of-sight velocity $v_z$, 
in a reference frame where 
the proton velocity is zero.  Results are shown for 
fixed $f_p=0.5$, several values of $\beta=0.01,0.1,0.5,1$, at shock velocities $v_s=1000$ km s$^{-1}$ (left panel) 
and $v_s=7000$ km s$^{-1}$.  At low shock velocity, charge transfer is extremely efficient, and the 
broad neutral distributions are very close to the thermal proton distribution, with the broad neutrals moving 
slightly slower than the ions (left panel).  
At high shock velocity, the broad 
neutral distribution function is skewed and offset from that of the protons, leading to asymmetric velocity 
profiles with the broad neutrals moving considerably faster than the ions (right panel).}
\label{fig:profileFO}
\end{figure}

Eqs.~(\ref{eq:phiFO}) and (\ref{eq:phiEW}) allow us to model the FWHM of the broad line as a function of $v_s$ and 
$\beta$, as shown in Fig.~\ref{fig:FWHM}.  
In the extreme scenarios of Case A and Case B conditions, the FWHM has a weak ($\la 1\%$) dependence on the 
pre-shock ionization after taking a spatial average over the shock transition zone.  
Figure~\ref{fig:profileFO} shows examples of broad neutral velocity distributions 
[which are inputs to eq.~(\ref{eq:phiFO})] for the case of face-on orientation 
of the shock as a function of line-of-sight velocity $v_r$, in a frame of reference where the ion velocity is 
zero.  Results are shown for fixed $f_p=0.5$ at several values of $\beta=0.01,0.1,0.5,1$ at shock velocities 
$v_s=1000$ km s$^{-1}$ (left panel) and $v_s=7000$ km s$^{-1}$ (right panel).  At low shock velocities, charge transfer is 
extremely efficient, yielding broad neutral distribution functions nearly identical to that of thermal protons, 
with the broad neutral moving slightly slower than the ions. 
At high shock velocities, the broad neutral distributions are skewed and offset from the proton distribution, 
leading to the asymmetric profiles depicted in the right panel, with the broad neutrals moving considerably faster 
than the ions.

We write the FWHM relation as a function of $v_s$ and $\beta$:
\begin{eqnarray}
\label{eq:FWHM}
W_{\rm FWHM} = W(v_s,\beta),
\end{eqnarray}
where $W_{\rm FWHM}$ and $W$ represent the measured and theoretical values of the line FWHM, respectively.
For SNRs with measured $W_{\rm FWHM}$ and $I_b/I_n$, combining eqs.~(\ref{eq:ibintheory}) and (\ref{eq:FWHM}) 
self-consistently 
constrains $v_s$ and $\beta$.  This is accomplished by inverting $W(v_s,\beta)$ to yield 
$v_s=W^{-1}(W_{\rm FWHM},\beta)$,
and calculating the root of the expression
\begin{eqnarray} 
L_{bn}\left[W^{-1}(W_{\rm FWHM},\beta),\beta\right]-I_b/I_n=0.
\end{eqnarray}
This procedure produces a pair of values 
$(v_s,\beta)$ for which the theoretical calculations $L_{bn}$ and $W$ equal the measured $I_b/I_n$ and $W_{\rm FWHM}$.

We emphasize that a self-consistent calculation is critical for accurately determining the 
values of $v_s$ and $\beta$.  In the previous literature, two bracketing values for 
the temperature equilibration are sometimes chosen (e.g., $\beta=0.1,1$), and the FWHM relation is employed 
to give a range of possible values for the shock velocity \citep[e.g.,][]{HengMcCray07a}.  
However, this procedure has a major flaw.  In practice, the self-consistent quantity 
$L_{bn}\left[W^{-1}(W_{\rm FWHM},\beta),\beta\right]$ will have a minimum value over the range $\beta\in (m_e/m_p,1)$.  
If $I_b/I_n$ for a particular observed shock is less than this minimum value, then no pair 
$(v_s,\beta)$ will yield the observed values $I_b/I_n$ and $W_{\rm FWHM}$.  
In such a case, the model breaks down and quoting a range of possible shock velocities for two bracketing values of 
$\beta$ is inappropriate.  Additional physics must be invoked to account for the observations.
Moreover, when no 
measurement of $I_b/I_n$ exists, the bracketing procedure may or may not yield an accurate estimate of the range of 
shock velocities.

We use the methodology described above to self-consistently extract shock parameters from observations of 
Balmer-dominated SNRs.
We summarize our results in Table~\ref{table:obs}, which lists, from left to right, the object, reference, 
H$\alpha$ $W_{\rm FWHM}$ and $I_b/I_n$, calculated values for $v_s$ and $\beta$ from H$\alpha$, Ly$\beta$ 
$W_{\rm FWHM}$, and calculated value for $v_s$ from Ly$\beta$.  
If our models do not yield a fit to the observations, we do not list values 
for the shock velocity and temperature equilibration ratio.  
Our calculations show a characteristic range of $\beta$ between $0.01$ and $0.1$ for $v_s>1000$ km s$^{-1}$.
Therefore, in the case of Ly$\beta$ observations where no measurement 
of $I_b/I_n$ exists, we report derived shock velocities for the range $\beta=0.01-0.1$; further 
observations are required to confirm the accuracy of these estimates.  For SNR 0519-69.0, observations 
exist in 
H$\alpha$ and Ly$\beta$.  To calculate shock velocities from the Ly$\beta$ observations, we use the derived value of 
$\beta$ from the H$\alpha$ diagnostics.

Our models successfully fit 14 measurements from seven Balmer-dominated SNRs, within the observational uncertainties.  
In the cases of the Cygnus Loop \citep[][]{Ghavamianetal01a} and one measurement from 0519-69.0 by
\citet[][]{Tuohyetal82a}, the observed ratio $I_b/I_n$ is too low to be accounted for by our calculations.  In 
addition, we are unable to fit the majority of measurements from the object DEM L71/0505-67.9, which we have omitted 
from the table.  Below, we discuss a possible explanation for these discrepancies in our model.

We note that our inferred shock velocities and temperature 
equilibration ratios are systematically lower than in previous studies, and show an increased sensitivity 
to $\beta$ compared to Papers 1 and 2.  
The ability to 
sensitively probe both shock velocity and temperature equilibration is of interest for studies of collisionless electron 
heating in shocks, as described below.
Typically, inferred values for $v_s$ are $\sim 10-30\%$ smaller than those quoted in Paper 1, 
primarily because of the contribution of broad neutral velocities to relative speeds in fast 
neutral-ion interactions.  The shock speeds in Tycho's SNR and SN1006 are smaller by about 15\% and 27\%, respectively, 
than those reported previously. 
One implication is that the distances to these SNRs derived from the shock speeds and proper motions are correspondingly 
smaller.  The inferred distance to SN1006 is reduced from 2.18 kpc \citep[][]{Winkleretal03a}, to 1.6 kpc.  The 
corresponding brightness of the SN is squarely in the middle of the Type I SN distribution at 2.18 kpc, 
but is 0.7 magnitudes fainter at 1.6 kpc.  The smaller distance is more comfortable in comparison with that of the 
S-M star which lies between 1.05~kpc and 2.1~kpc and whose spectrum shows absorption by SN1006 ejecta 
\citep[][]{Burleighetal00a}.  On the other hand, the ejecta are observed to expand at 7026 km s$^{-1}$ 
\citep[][]{Hamiltonetal07a}, and the requirement that this material lies within the remnant places 
a conservative lower limit to the distance of 1.6 kpc, which is just consistent with that derived here.

\section{Discussion}
\label{sect:Discussion}

\subsection{Collisionless Shock Heating}
\label{subsect:Collisionless}

One of the motivations for studying Balmer-dominated SNRs is to probe the physics of collisionless 
shocks.  While various mechanisms for transferring energy from proton to electron populations have been proposed in 
the literature, there is no consensus on how to predict the value 
of $\beta$ for a given set of 
shock parameters in astrophysical contexts.  In a recent paper, \citet[][]{Ghavamianetal07b} attempted to address this 
problem by deriving a relation for temperature equilibration ratio versus shock velocity from observations of shocks 
in SNRs.  They reported that their results could be fit by a curve $\beta(v_s)\propto v_s^{-2}$, with $\beta=1$ for 
$v_s\le 400$ km s$^{-1}$.  
Given that the proton temperature scales roughly as $T_p\propto v_s^2$, this conclusion implies that the electrons are 
heated to a constant temperature, independently of the shock velocity.  While \citet[][]{Ghavamianetal07b} discussed a 
possible mechanism for this dependence, it has not yet been established that theoretical models can produce such an 
effect.

\begin{figure}
\begin{center}
\plotone{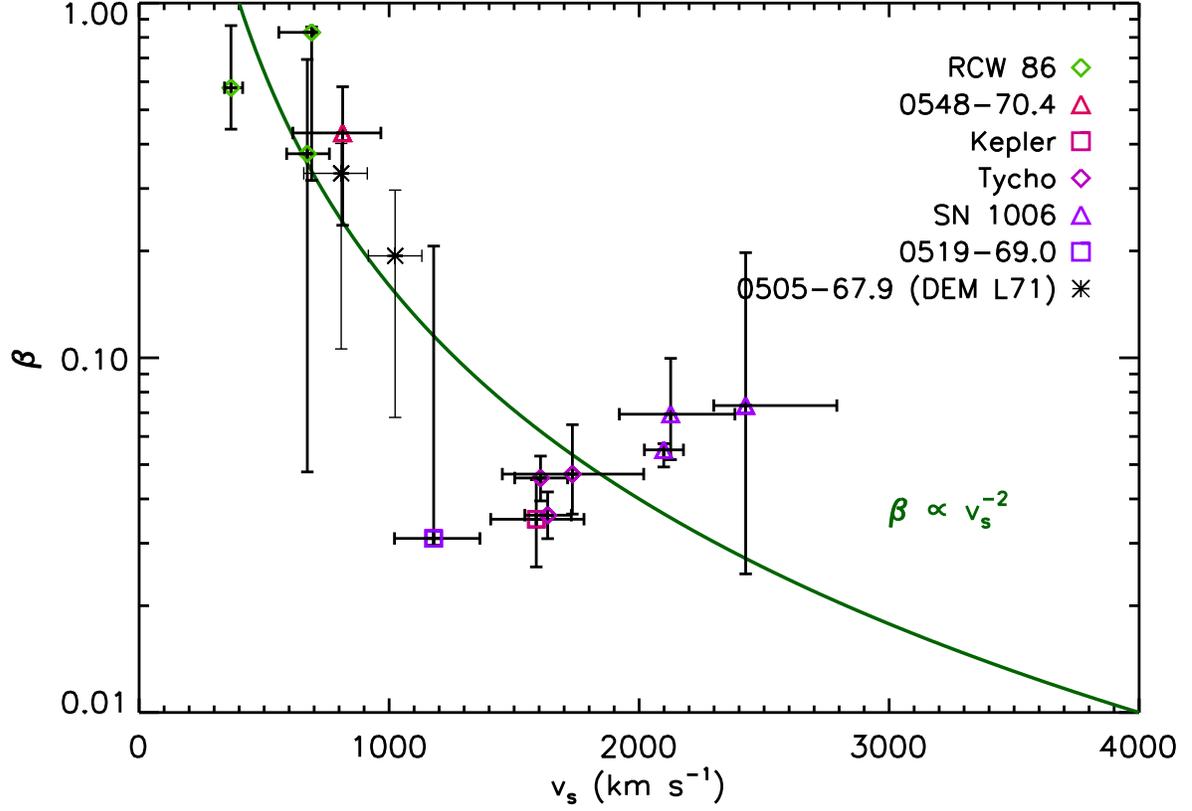}
\end{center}
\caption{Temperature equilibration ratio $\beta$ as a function of shock velocity $v_s$ for Balmer-dominated shocks 
fit by our calculations.  The symbol shape denotes the origin of the data point from SNRs in our sample.  
The solid curve shows the dependence $\beta(v_s)\propto v_s^{-2}$ for a heating mechanism which is independent of 
shock velocity.  Fitting this curve to the inferred values from our new model yields a reduced chi-square of 
$\chi_r^2 = 62.8/12 = 5.2$.}
\label{fig:betacurve}
\end{figure}

Using our new model for emission from Balmer-dominated shocks, we have calculated an updated $\beta(v_s)$ relation, 
which is displayed in Figure~\ref{fig:betacurve}.  
The solid curve depicts the proposed $v_s^{-2}$ dependence, with $\beta=1$ for $v_s<400$ km s$^{-1}$.
For all the points shown in the plot, we have 
self-consistently fit measurements of both $W_{\rm FWHM}$ and $I_b/I_n$.   
We fix the pre-shock ionization at $f_p=0.5$,  
and exclude the majority of measurements from DEM L71/0505-67.9, as well as one measurement each from Cygnus and 
0519-69.0 which we are unable to account for with our models.  
We find from our calculations that the data are not well fit by a 
power-law relation $\beta(v_s)\propto v_s^{-\alpha}$; if we set $\alpha=2$, the fit to the $\beta$ curve yields a 
reduced chi-square of $\chi_r^2=62.8/12=5.2$.\footnote{Note that such a large value of $\chi_r^2$ should be interpreted 
in the context of $10\%$ uncertainties associated with the atomic cross sections used.}
For shock velocities $v_s\la 1500$ km s$^{-1}$, our results are fairly close to those of previous models.  At 
higher velocities $v_s> 2000$ km s$^{-1}$, the deviations are more significant.
The minimum 
value of the temperature equilibration ratio is $\beta\sim 0.03$ at velocities $v_s\ga 1500$~km~s$^{-1}$.  This 
value is greater by several orders of magnitude than the theoretical minimum $\beta=m_e/m_p$, 
but is smaller than the value predicted by some collisionless heating models
\citep[e.g.,][]{CargillPapadopoulos88a}.

\subsection{Limitations to our Model and Future Work}
\label{subsect:Limitations}

As demonstrated by SNR 0505-67.9, the Cygnus Loop, and 0519-69.0, an additional physical mechanism is needed 
to account for the observed 
H$\alpha$ emission seen in some SNRs. 
One possibility is that a significant amount of the dissipated energy in the shock is transferred to cosmic rays, 
producing a precursor which can heat and accelerate 
the upstream gas, altering the shock jump conditions   
\citep[][]{Smithetal91a,Hesteretal94a,Sollermanetal03a}.  
In addition, the precursor can ``push'' on the upstream protons, leading to a velocity 
differential between neutrals and charged species \citep[e.g.,][]{BerezhkoEllison99a}.  
Furthermore, all previous models have assumed that the kinetic 
distribution functions for protons and electrons in the shock transition zone are Maxwellian.  However, recent 
calculations have shown that the proton distribution can significantly depart from Maxwellian behavior, forming  
a distinct ``pickup'' ion population \citep[][]{Raymondetal08a}. 
This effect will change the structure of the transition zone, broad neutral distributions, and kinetic reaction rates. 

The inclusion of a cosmic ray precursor will affect the predicted 
broad-to-narrow line strength in several ways.  
Broadening of the cold neutral distribution function effectively reduces the 
optical depth to Ly$\beta$ scattering, which will decrease the conversion efficiency of narrow Ly$\beta$ to H$\alpha$ 
and increase $L_{bn}$.  In contrast, additional excitation of cold neutral H atoms in the precursor will increase 
narrow line emission and decrease $L_{bn}$.  While it is not obvious which effect is dominant, 
evidence for such precursor effects exists in 
the anomalously large widths of H$\alpha$ lines in most Balmer-dominated SNRs \citep[][]{Sollermanetal03a}.
In addition, H$\alpha$ emission from a spatially resolved precursor in Tycho's SNR has recently been reported by 
\citet[][]{Leeetal07a}.
 
The spatial structure of the shock transition zone provides a way to infer the external density $n_0$ 
and pre-shock 
ionization fraction $f_p$.  As noted in \S\ref{subsect:EmissivityResults} the spatial emissivity profile and centroid 
shift between the narrow 
and broad components have a strong dependence on $f_p$ and $n_0$.  
\citet[][]{Raymondetal07a} were able to spatially resolve the total H$\alpha$ emission from SN1006 using the 
ACS camera on the {\it Hubble Space Telescope}.  
\citet[][]{Raymondetal07a} adopted the values of $v_s$ and $\beta$ from those inferred by \citet[][]{Ghavamianetal02a} 
using observations of the FWHM and $I_b/I_n$, and then attempted to fit for the spatial structure by varying 
$f_p$ and $n_0$.  Future instruments similar to ACS with high spatial resolution (e.g., WFC3 onboard {\it Hubble} 
with the narrow H$\alpha$ component isolated by a narrow band filter) will allow for more accurate measurements of the 
transition zone structure and provide additional constraints on shock parameters.

\begin{figure}
\begin{center}
\plotone{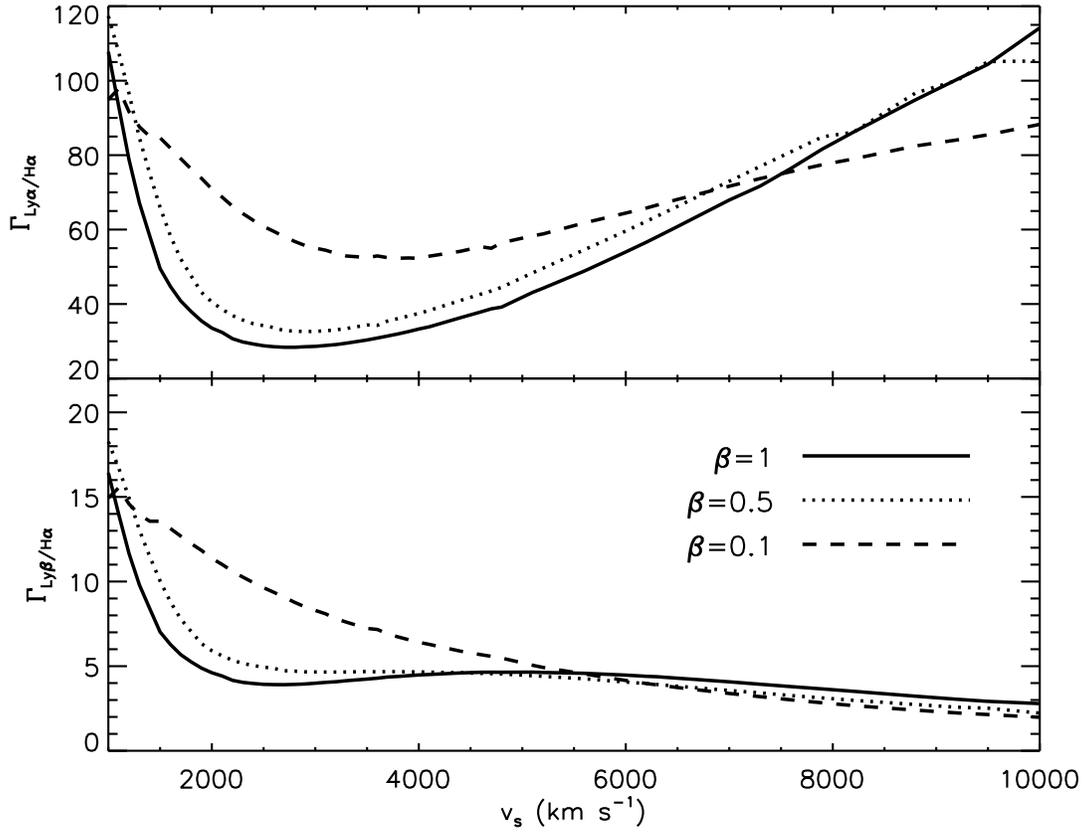}
\end{center}
\caption{Luminosity ratios of broad Ly$\alpha$ and Ly$\beta$ to H$\alpha$, denoted $\Gamma_{Ly\alpha/H\alpha}$ and 
$\Gamma_{Ly\beta/H\alpha}$, respectively.}
\label{fig:LyHratios}
\end{figure}

The models presented in our series of papers generically describe the physics of non-radiative shocks interacting 
with cold, pre-shock gas.  In addition to using them 
to understand nearby SNRs, we can also calculate hydrogen line emission from SNRs in young, distant galaxies 
(with redshifts $3-5$), a subject first explored by \citet[][]{HengSunyaev08a}.  We improve on their estimates of 
the luminosity ratios of Ly$\alpha$ and Ly$\beta$ to H$\alpha$, denoted $\Gamma_{Ly\alpha/H\alpha}$ and 
$\Gamma_{Ly\beta/H\alpha}$, respectively (see Figure~\ref{fig:LyHratios}; c.f. Figure~1 of \citealt[][]{HengSunyaev08a}).
\citet[][]{HengSunyaev08a} underestimated the broad Ly$\alpha$ emission at high shock velocities due to their 
neglect of the multi-component shock transition zone.  Nevertheless, their conclusions remain intact: 
the sensitivity of $\Gamma_{Ly\alpha/H\alpha}$ and $\Gamma_{Ly\beta/H\alpha}$ to $\beta$ by a factor $\sim 2$ over 
the velocity range $1000\la v_s \la 4000$ km s$^{-1}$ is a direct and unique way to measure the temperature ratio.  A 
valuable extension to their work will incorporate models of Ly$\alpha$ and Ly$\beta$ scattering, taking into account 
geometric effects.

The biggest systematic uncertainties in our current model are: i) a proper treatment of Lyman line scattering in the 
shock transition zone; ii) the effect of a cosmic ray precursor on $L_{bn}$; and iii) inclusion of 
a non-thermal population of protons in calculating the reaction kinetics.  The incorporation of these physical effects is 
important for future models.  Our new results yield substantial differences in derived shock speed 
(and hence inferred distances) for measurements of Balmer-dominated SNRs compared to previous models.  While our results 
are not consistent with a power-law relation between shock velocity and electron temperature, future work is needed to 
resolve the remaining approximations in our model and make a more definitive statement about shock heating of electrons.

\acknowledgements

We thank Carles Badenes, Anatoly Spitkovsky and Eliot Quataert for useful conversations, and Parviz Ghavamian for 
helping clarify several points in the observational literature.
K.H. thanks the Institute for Advanced Study for their generous support.

\appendix

\section{Shock Kinetics}
\label{sect:AppendixA}

We summarize important details of our calculations of rate coefficients, broad neutral velocity distributions and 
broad line profiles.  For a more detailed description of our methods, see Paper 1.  In this work, we treat the 
$nl$ atomic sub-levels 
separately instead of considering a single, summed $n$ level as in Paper 1.  We treat charge transfers, excitation 
and ionization events between electrons, protons, alpha particles and hydrogen atoms using the cross sections of 
\citet[][]{Barnettetal90a}, \citet[][]{Belkicetal92a}, \citet[][]{JanevSmith93a}, \citet[][]{Balancaetal98a}, 
and \citet[][]{Hareletal98a}.  Fitting functions for some of these cross sections are provided in 
\citet[][]{HengSunyaev08a}.

\begin{figure}
\begin{center}
\plotone{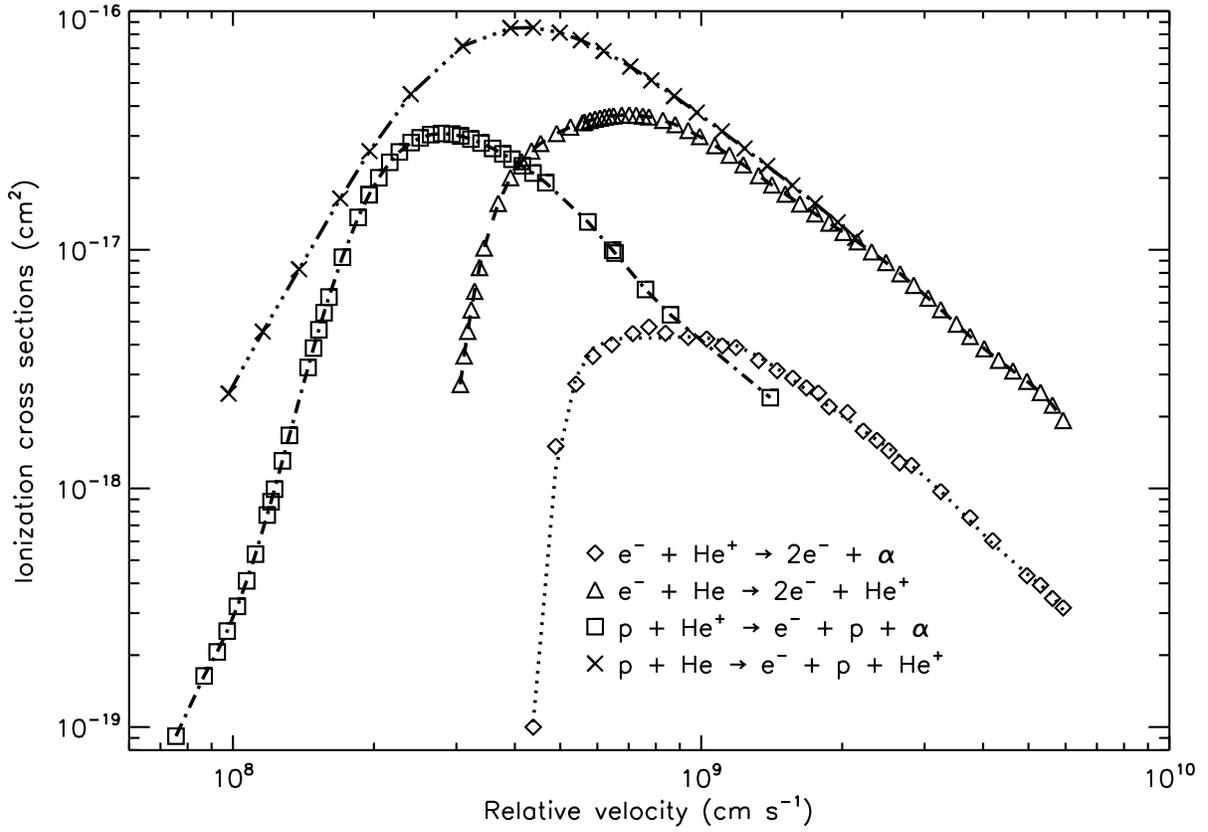}
\end{center}
\caption{Cross sections for ionization of neutral and singly-ionized helium by electrons and protons.  For references to 
the atomic data, see Appendix~\ref{sect:AppendixA}.}
\label{fig:crosssections}
\end{figure}

We also consider the ionization of helium atoms and singly-ionized helium by electrons and protons, using the 
cross sections of \citet[][]{Peartetal69a}, \citet[][]{Angeletal78a}, \citet[][]{Ruddetal83a}, 
\citet[][]{ShahGilbody85a}, \citet[][]{Rinnetal86a} and \citet[][]{Shahetal88a}.  
Following \citet[][]{Barnettetal90a}, we fit the cross sections to the function
\begin{eqnarray}
{\mathcal F}\left(x;{\bf A}\right) = \exp{\left({A_0\over 2}{\mathcal C}_0 + \sum^{8}_{i=1} ~A_i 
	~{\mathcal C}_i\left(x\right)\right)},
\end{eqnarray}
where the components of ${\bf A}=(A_0,A_2,\ldots,A_8)$ are the fitting parameters, and the  
quantities ${\mathcal C}_i$ are the Chebyshev polynomials:
\begin{eqnarray}
{\mathcal C}_0\left(x\right) &=& 1,\\
{\mathcal C}_1\left(x\right) &=& x,\\
{\mathcal C}_i\left(x\right) &=& 2x{\mathcal C}_{i-1} - C_{i-2}.
\end{eqnarray}
We define the fitting variable $x$ as
\begin{eqnarray}
x= { \ln{\left[E^2/(E_{\rm max} E_{\rm{min}})\right]} 
	\over \ln{\left(E_{\rm{max}}/E_{\rm{min}}\right)}},
\end{eqnarray}
where $E$ is the relative energy between the interacting particles and $E_{\rm min}$ and $E_{\rm max}$ are the 
minimum and maximum energies for which data are available.  We assume a fiducial error of 10$\%$ for the data.  
The cross sections and corresponding fits are displayed in Figure~\ref{fig:crosssections} and the fitting parameters 
are presented in Table~\ref{table:fitparams}.

We neglect charge transfer reactions of neutral and singly-ionized helium with protons due to scarcity of 
cross sections for these processes in our velocity range of interest.  While such interactions may affect helium 
line emission, we do not expect them to strongly couple to the Balmer radiation.  In addition, for ease of 
computation, we approximate 
excitation and ionization of hydrogen by He$^+$ using relevant proton rate coefficients.\footnote{Our assumption 
is based on the Weizsacker-Williams 
approximation in which scattering dynamics are dominated by the charge of the impacting particle as opposed to its mass 
\citep[e.g.,][]{Jackson98a}.  This approximation is valid when the relative velocity of the collision is greater than 
that of an electron orbiting the hydrogen atom.  At low velocities $v_s\lesssim 300$ km s$^{-1}$, this assumption breaks 
down.}
However, cross sections for this process do exist in the literature and should be used in future 
calculations \citep[][]{Barnettetal90a}.  Nevertheless, we expect corrections to these cross sections to have a 
small quantitative effect on our results.

We make several approximations to speed up our computations.  For temperature ratios $0.01\la\beta\la 0.1$, the 
velocity width of the electron distribution is generally broader than that of the broad neutrals (characteristic 
electron velocities are greater than proton velocities by a factor of $4-14$), implying that rate coefficients for 
interactions involving electrons are insensitive to changes in the broad neutral velocity distribution.  We therefore 
approximate electron rate coefficients to be the same for reactions involving cold atoms and broad neutrals.  
Calculations of the FWHM for line profiles should include excitations by electrons, protons, singly-ionized helium and 
alpha particles.  However, alpha particles contribute $\la 1\%$ due to the relatively smaller density $n_{\alpha}$, 
and can typically be omitted. 

The reaction rate coefficients used in eqs.~(\ref{eq:MassFlux1})--(\ref{eq:MassFlux6}) are defined as:
\begin{eqnarray}
\label{eq:Rdef}
\tilde{R}_{X,s} = \int\, d^3{\bf v}\, \int\, d^3{\bf v}'\, f_a({\bf v})\, f_s({\bf v}')\, \Delta v\, 
	\sigma_{X,s}(\Delta v),
\end{eqnarray}
where $X$ denotes the interaction (ionization or charge transfer), $a$ the atomic species (broad or cold neutrals), 
and $s$ the interacting charged particle (electrons, protons, singly-ionized helium, or alpha particles).  In contrast 
to eqs.~(\ref{eq:phiEW})--(\ref{eq:phiFr}), the interaction cross sections $\sigma_{X,s}$ used here 
represent the sum of the reactions to all $nl$ levels for which atomic data is available 
[c.f., eq.~(\ref{eq:sigHalpha})].  
Using the definition of eq.~(\ref{eq:Rdef}), the quantity $n_a n_s \tilde{R}_{X,s}$ gives the number of interactions $X$, 
between the species $a$ and $s$, per unit volume.

\section{Typographical Errors in Papers 1 \& 2}

We point out several minor typographical errors in \citet[][]{HengMcCray07a} and \citet[][]{Hengetal07a}.  
In \S5.1 of \citet[][]{HengMcCray07a}, ${\mathcal R}_{bn}(\rm{H}\alpha) = I_b(\rm{H}\alpha)/I_b(\rm{H}\alpha)$ should 
be ${\mathcal R}_{bn}(\rm{H}\alpha) = I_b(\rm{H}\alpha)/I_n(\rm{H}\alpha)$.  
In \S5 of \citet[][]{Hengetal07a}, $R_{\rm{H}\alpha b_0}$ 
should be $R_{\rm{H}\alpha,b0}$, $R_{\rm{H}\alpha b_*}$ should be $R_{\rm{H}\alpha,b_*}$ and $R_{\rm{H}\alpha n}$ 
should be $R_{\rm{H}\alpha,n}$.  In \S6.2, paragraph 2 of \citet[][]{Hengetal07a}, the sentence that begins 
``Before CKR80 and HM07, \ldots'' should be changed to ``Before (CKR80 and HM07), \ldots''.  

\vfill\eject

\bibliographystyle{apj}
\bibliography{ms}

\begin{landscape}
\begin{table}
\begin{center}
\caption{Inferred Shock Velocities and Temperature Equilibrations from SNRs}
\label{table:obs}
\begin{tabular}{lccccccccc}
\tableline
\multicolumn{1}{c}{Object} & \multicolumn{1}{c}{Reference} & \multicolumn{1}{c}{H$\alpha$ FWHM} & \multicolumn{1}{c}{H$\alpha$ $I_b/I_n$} & \multicolumn{1}{c}{H$\alpha$ $v_s$} & \multicolumn{1}{c}{$\beta$} & \multicolumn{1}{c}{Ly$\beta$ FWHM} & \multicolumn{1}{c}{Ly$\beta$}\\
\multicolumn{1}{c}{} & \multicolumn{1}{c}{} & \multicolumn{1}{c}{(km s$^{-1}$)} & \multicolumn{1}{c}{} & \multicolumn{1}{c}{(km s$^{-1}$)} & \multicolumn{1}{c}{} & \multicolumn{1}{c}{(km s$^{-1}$)} & \multicolumn{1}{c}{(km s$^{-1}$)}\\
\tableline
Cygnus & G01 & $262 \pm 32$ & $0.59 \pm 0.3$ & --- & --- & --- & ---\\
RCW 86 & G07b & $325 \pm 10$ & $1.06 \pm 0.1$ & $367^{+46}_{-26}$ & $0.577^{+0.287}_{-0.136}$ & --- & ---\\
 & G01 & $562 \pm 18$ & $1.18 \pm 0.03$ & $690^{+9}_{-131}$ & $0.826^{+0.030}_{-0.510}$ & --- & ---\\
 & G07b & $640 \pm 35$ & $1.0 \pm 0.2$ & $673^{+89}_{-83}$ & $0.376^{+0.317}_{-0.328}$ & --- & ---\\
0505---67.9 & S91 & $580 \pm 70$ & $\gtrsim 0.7$ & --- & --- & --- & ---\\
 & G03 & $785^{+95}_{-80}$ & $0.93^{+0.11}_{-0.10}$  & $808^{+104}_{-150}$ & $0.331^{+0.071}_{-0.225}$ & --- & ---\\
 & G03 & $1055^{+150}_{-120}$ & $0.88^{+0.11}_{-0.10}$  & $1024 \pm 107$ & $0.194^{+0.103}_{-0.126}$ & --- & ---\\
 & G07a & --- & ---  & --- & --- & $1135 \pm 30$ & $1036 \pm 46$$^*$\\
 & G07a & --- & ---  & --- & --- & $1365 \pm 75$ & $1249 \pm 89$$^*$\\
0548---70.4 & S91 & $760 \pm 140$ & $1.1 \pm 0.2$ & $814^{+153}_{-199}$ & $0.430^{+0.150}_{-0.194}$ & --- & ---\\
0519---69.0 & S91 & $1300 \pm 200$ & $0.8 \pm 0.2$ & $1178^{+185}_{-157}$ & $0.031^{+0.176}_{-0.001}$ & --- & ---\\
 & T82 & $2800 \pm 300$ & 0.4---0.8$^\dagger$  & --- & --- & --- & ---\\
 & G07a & --- & --- & --- & --- & $3130 \pm 155$ & $2984^{+703}_{-185}$ \\
Kepler & F89 & $1750 \pm 200$ & $1.1 \pm 0.25$ & $1589^{+191}_{-182}$ & $0.035^{+0.010}_{-0.009}$ & --- & ---\\
Tycho & G01 & $1765 \pm 110$ & $0.67 \pm 0.1$ & $1606^{+108}_{-103}$ & $0.046^{+0.007}_{-0.006}$ & --- & ---\\
 & KWC87 & $1800 \pm 100$ & $1.08 \pm 0.16$ & $1634^{+95}_{-91}$ & $0.036^{+0.006}_{-0.005}$ & --- & ---\\
 & S91 & $1900 \pm 300$ & $0.77 \pm 0.09$ & $1733^{+285}_{-280}$ & $0.047^{+0.018}_{-0.011}$ & --- & ---\\
SN 1006 & G02 & $2290 \pm 80$ & $0.84^{+0.03}_{-0.01}$ & $2098^{+79}_{-77}$ & $0.055^{+0.002}_{-0.006}$ & --- & ---\\
 & S91 & $2310 \pm 210$ & $0.73 \pm 0.06$ & $2126^{+257}_{-205}$ & $0.069^{+0.030}_{-0.018}$ & --- & ---\\
 & KWC87 & $2600 \pm 100$ & $0.77 \pm 0.08$ & $2426^{+364}_{-128}$ & $0.073^{+0.125}_{-0.049}$ & --- & ---\\
\tableline
\end{tabular}
\end{center}
\scriptsize
Note: Case A conditions are assumed for the broad line.  The pre-shock ionization fraction is taken to be $f_p=0.5$.\\
$*$: Values are reported for the range $0.01 \le \beta \le 0.1$.\\
$\dagger$: We regard this as $0.6 \pm 0.2$ when solving for $\beta$ and $v_s$.\\
\normalsize
\end{table}
\end{landscape}

\begin{table}
\begin{center}
\caption{Fits to Helium Species Ionization Cross Sections}
\label{table:fitparams}
\begin{tabular}{lcccc}
\tableline\tableline
\multicolumn{1}{c}{} & \multicolumn{1}{c}{He + e$^{-}$} & \multicolumn{1}{c}{He$^{+}$ + e$^{-}$} & \multicolumn{1}{c}{He + p} & \multicolumn{1}{c}{He$^{+}$ + p}\\
\tableline
$A_0$ & -78.4712 & -82.6155 & -77.0261 & -80.7740\\
$A_1$ & -0.832236 & -0.535745 & 0.596233 & 1.50686\\
$A_2$ & -1.00452 & -1.15893 & -1.37165 & -2.04982\\
$A_3$ & 0.482606 & 0.644513 & 0.205854 & 0.384422\\
$A_4$ & -0.244927 & -0.419765 & 0.123038 & 0.353731\\
$A_5$ & 0.121965 & 0.299795 & -0.0671246 & -0.268015\\
$A_6$ & -0.0795100 & -0.204216 & -0.0135306 & -0.0107265\\
$A_7$ & 0.0537985 & 0.133238 & 0.0163305 & 0.0101069\\
$A_8$ & -0.0521662 & -0.0716388 & -0.00556328 & -0.110770\\
$E_{\rm min}$ & 26.6 eV & 54.5 eV & 5.0 keV & 2.98 keV\\
$E_{\rm max}$ & $10^4$ eV & $10^4$ eV & $2.38 \times 10^3$ keV & $1.03\times 10^3$ keV\\
\tableline
\end{tabular}
\end{center}
\end{table}

\end{document}